\documentclass{aa}
\usepackage{psfig,graphicx,ulem,epsfig}
\usepackage{txfonts}
\def\oph{Ophiuchus}

\begin{document}

\title{Ophiuchus: an optical view of a very massive cluster of
  galaxies hidden behind the Milky Way \thanks{Based on observations
    obtained with MegaPrime/MegaCam (program 10AF02), a joint project
    of CFHT and CEA/DAPNIA, at the Canada-France-Hawaii Telescope
    (CFHT) which is operated by the National Research Council (NRC) of
    Canada, the Institute National des Sciences de l'Univers of the
    Centre National de la Recherche Scientifique of France, and the
    University of Hawaii. Based on observations made with ESO
    Telescopes at the La Silla Paranal Observatory under programme ID
    085.A-0016(C). Based on observations obtained at the Southern
    Astrophysical Research (SOAR) telescope (program 2009B-0340 on
    SOI/SOAR), which is a joint project of the Minist\'erio da
    Ci\^encia, Tecnologia, e Inova\c c\~ao (MCTI) da Rep\'ublica
    Federativa do Brasil, the U.S. National Optical Astronomy
    Observatory (NOAO), the University of North Carolina at Chapel
    Hill (UNC), and Michigan State University (MSU). This research has
    made use of the NASA/IPAC Extragalactic Database (NED) which is
    operated by the Jet Propulsion Laboratory, California Institute of
    Technology, under contract with the National Aeronautics and Space
    Administration, and of the SIMBAD database, operated at CDS,
    Strasbourg, France.}}

\author{
F.~Durret \inst{1} \and
K.~Wakamatsu \inst{2} \and
T.~Nagayama \inst{3} \and
C.~Adami \inst{4} \and
A.~Biviano \inst{5,1} 
}

\institute{
Sorbonne Universit\'es, UPMC Univ. Paris 6 et CNRS, UMR~7095, Institut d'Astrophysique de Paris, 98bis Bd Arago, 75014, Paris, France
\and
Faculty of Engineering, Gifu University, 1-1 Yanagido, Gifu 501-1193, Japan
\and
Department of Astrophysics, Nagoya University, Furocho, Chikusaku, 
Nagoya 464-8602, Japan
\and
LAM, OAMP, P\^ole de l'Etoile Site Ch\^ateau-Gombert, 38 rue Fr\'ed\'eric 
Joliot--Curie,  13388 Marseille Cedex 13, France
\and
INAF/Osservatorio Astronomico di Trieste, via Tiepolo 11, 34143 Trieste, Italy
}

\date{Accepted . Received ; Draft printed: \today}

\authorrunning{Durret et al.}

\titlerunning{Optical view of the Ophiuchus cluster}

\abstract
{The Ophiuchus cluster, at a redshift z=0.0296, is known from X-rays
  to be one of the most massive nearby clusters, but due to its very
  low Galactic latitude its optical properties have not been
  investigated in detail.}
{We discuss the optical properties of the galaxies in the Ophiuchus cluster,
in particular with the aim of understanding better its dynamical properties.}
{We have obtained deep optical imaging in several bands with various
  telescopes, and applied a sophisticated method to model and subtract
  the contributions of stars in order to measure galaxy magnitudes as
  accurately as possible. The colour-magnitude relations obtained show
  that there are hardly any blue galaxies in \oph\ (at least brighter
  than $r'\leq 19.5$), and this is confirmed by the fact that we only
  detect two galaxies in H$\alpha$. We also obtained a number of
  spectra with ESO-FORS2, that we combined with previously available
  redshifts. Altogether, we have 152 galaxies with spectroscopic
  redshifts in the $0.02 \leq z \leq 0.04$ range, and 89 galaxies with
  both a redshift within the cluster redshift range and a measured
  $r'$ band magnitude (limited to the Megacam $1\times 1$~deg$^2$
  field). }
{A complete dynamical analysis based on the galaxy redshifts available
  shows that the overall cluster is relaxed and has a mass of
  1.1~$10^{15}$~M$_\odot$. The Sernal--Gerbal method detects a main
  structure and a much smaller substructure that are not separated in
  projection.}
{From its dynamical properties derived from optical data, the \oph\
  cluster seems to be overall a relaxed structure, or at most a minor
  merger, though in X-rays the central region (radius $\sim 150$~kpc)
  may show evidence for merging effects.}


\maketitle

\section{Introduction}

\begin{figure*} 
\centering \mbox{\psfig{figure=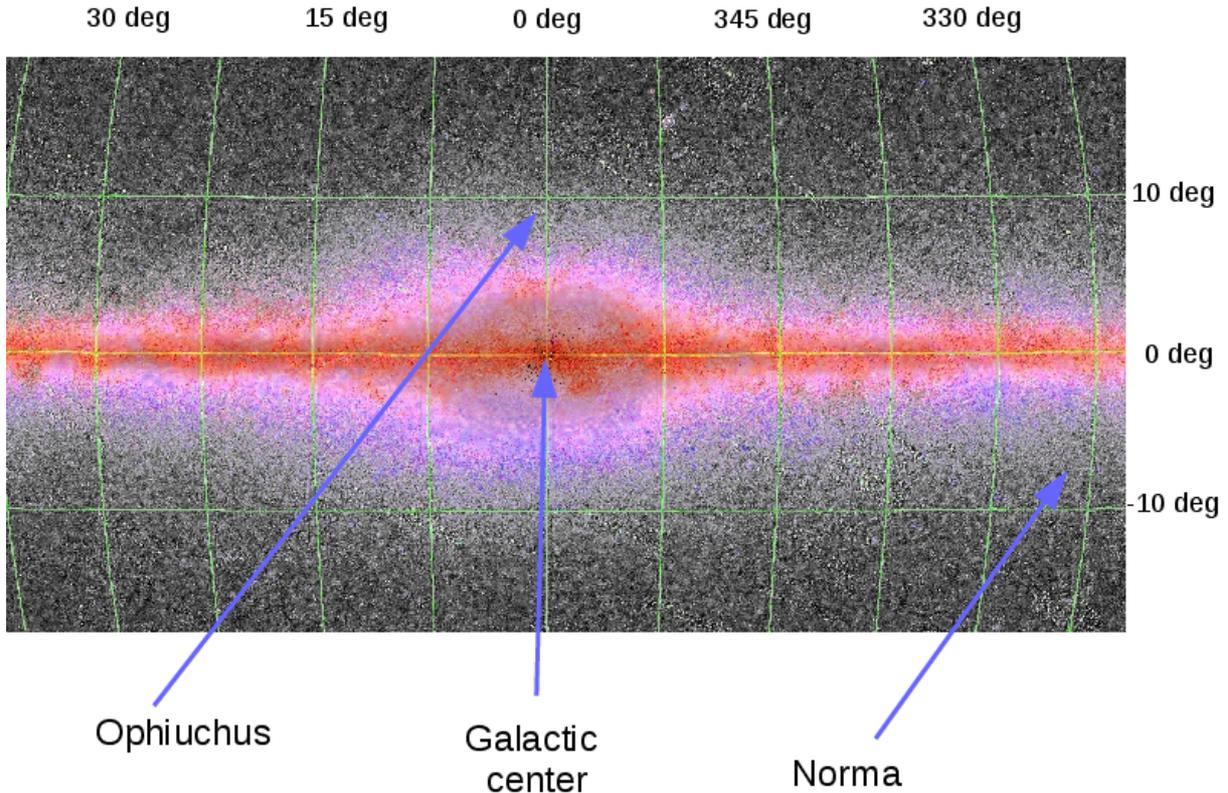,width=16cm}}
\caption{2MASS near-infrared image of the Milky Way taken from the
  Aladin database, showing the positions of the Galactic centre and of
  the \oph\ and Norma clusters.}
\label{fig:milkyway}
\end{figure*}

Though the study of clusters is presently more often devoted to large
surveys than to individual objects, it remains useful to analyse
clusters individually when they appear to have particularly
interesting or uncommon properties. Our attention was drawn on the
\oph\ cluster, that is the cluster with the second brightest X-ray
flux, but for which very little is known at optical wavelengths, due
to its very low galactic latitude.

Successful attempts have been made by several teams to detect
extragalactic objects behind the Milky Way. First, Kraan-Korteweg
(1989) searched for individual galaxies in the zone of avoidance at
optical wavelengths, and subsequently published a series of papers
with more and more detections, including clusters such as the Norma
cluster (Kraan--Korteweg et al. 1996, Skelton et al. 2009).  Nagayama
et al. (2006) performed a near-infrared study of CIZA~J1324.7-5736,
the second richest cluster of galaxies in the Great Attractor.  In
X-rays, a systematic search for clusters (and even superclusters) of
galaxies was made by Ebeling et al. (2002); in this paper, they give a
first catalogue of 73 clusters at redshifts $z<0.26$ and discuss the
identification of the Great Attractor. Other X-ray detections were
made by Lopes de Oliveira et al. (2006), who discovered Cl~2334+48 at
z=0.271 in the Zone of Avoidance in the XMM-Newton archive, and by
Mori et al. (2013), who detected the rich cluster Suzaku~J1759-3450 at
z=0.13 behind the Milky Way bulge. The latter cluster was then
confirmed by Coldwell et al. (2014), based on infrared data.

At a redshift of 0.0296 (as inferred from our dynamical analysis, see
Sect.~4.2), the \oph\ cluster was discovered in X-rays by Johnston et
al. (1981), who identified the 4U~$1708-23$ X-ray source with a
cluster of galaxies, and more or less simultaneously at optical
wavelengths by Wakamatsu \& Malkan (1981) during a search for highly
absorbed Galactic globular clusters.  It is located in the Zone of
Avoidance, not very far from the Great Attractor.

\oph\ is mostly known from its X-ray properties, since it is the
cluster with the second highest X-ray flux after Perseus (Edge et
al. 1990). The first results obtained with ASCA by Matsuzawa et
al. (1996) gave for the temperature and metallicity of the X-ray gas
the respective values of kT=9.8 keV and Z=0.24~Z$_\odot$.  Based on a
mosaic of ASCA data, Watanabe et al. (2001) later derived that \oph\
was formed by the merging of two clusters with different iron
abundances. They also detected the presence of a small group of
galaxies with a colder temperature superimposed on the main
cluster. \oph\ is part of the sample of clusters detected by Ebeling
et al. (2002) in their systematic X-ray search for clusters behind the
Milky Way. Ascasibar \& Markevitch (2006) showed that the X-ray
  emission map of \oph\ showed the presence of several sharp edges
  (see their Fig.~1), and argued that these could arise from gas
  sloshing, set up by a minor merger (as explained in detail by
  Markevitch \& Vikhlinin 2007).  The X-ray properties of the central
part of \oph\ were also analysed in detail by Million et al. (2010),
based on Chandra data. These authors showed the existence of a small
displacement between the X-ray peak and the cD galaxy ($\sim 2$~kpc),
and of several strong features, such as sharp fronts and a very steep
temperature gradient between the cool core (kT=0.7~keV) and a region
located only 30~kpc away (kT=10~keV). All these properties suggest
that the central regions of \oph\ (within $150$~kpc of the cluster
centre) show evidence for merging, whereas at radii larger than
150~kpc the cluster appears relatively isothermal with a constant
metallicity. At even higher energies, \oph\ is also part of the sample
of nearby clusters where evidence for indirect detection of dark
matter has been searched for in Fermi gamma-ray data (Hektor et
al. 2013).

Since \oph\ is very massive and very old (since it is in the nearby
Universe), it can be considered as a perfect example of a cluster
where the influence of the cluster on the member galaxies is as strong
as can be in the Universe, and it can be taken as a reference for
studies of galaxies in clusters such as red sequence or luminosity
functions.  \oph\ is also expected to be one of the clusters with very
little star formation, and this motivated our attempt to detect star
formation in \oph\ by observing it in a narrow band filter containing
the H$\alpha$ line at the cluster redshift. However, due to its low
Galactic latitude it has not been observed much at optical
wavelengths. An analysis was made of the galaxy distribution at very
large scale (covering a $12^\circ \times 17^\circ$ area) by Hasegawa et
al. (2000), based on 4021 galaxies with measured spectroscopic
redshifts.  It was then shown by Wakamatsu et al. (2005) that \oph\
has a large velocity dispersion of $1050\pm 50$~km~s$^{-1}$, agreeing
with its high X-ray luminosity, and that several groups or clusters
are located within $8^\circ$ of the cluster, thus forming a structure
comparable to a supercluster, close to the Great Attractor. They also
found a large foreground void, implying that it is a continuation of
the Local Void.

The difficulty of observing \oph\ at optical wavelengths is
illustrated in Fig.~\ref{fig:milkyway}, where we can see that it is
even closer to the Galactic centre than the Norma cluster.

For a redshift of 0.0296, Ned Wright's cosmology
calculator\footnote{http://www.astro.ucla.edu/~wright/CosmoCalc.html}
gives a luminosity distance of 120.8~Mpc and a spatial scale of
0.585~kpc/arcsec, leading to a distance modulus of 35.54 (assuming a flat
$\Lambda$CDM cosmology with H$_0=71$~km~s$^{-1}$~Mpc$^{-1}$, $\Omega
_M=0.27$ and $\Omega _\Lambda =0.73$). The cluster centre is
taken to be the position of the cD galaxy: RA=$258.1155^\circ$, Dec=$-23.3698^\circ$
(J2000.0). This is practically coincident with the peak of the X-ray
emission, which is $<3$ kpc (0.07 arcmin) away (Million et al. 2010).

The paper is organised as follows. We describe our optical data in
Section 2. Results concerning the red sequence detected in
colour-magnitude diagrams, star formation, internal structure and
dynamical properties are presented in Section 3.  The cluster merging
state is discussed in Section~4.

\section{Optical imaging and spectroscopy}
\label{sec:obs}

\subsection{The data}

\begin{table*}[t!]
\caption{Optical imaging observations. The field and pixel size are the
values of the reduced dithered and binned images.}
\begin{center}
\begin{tabular}{lccrccl}
\hline
\hline
          &            &        &           &       & & \\
Telescope & Instrument & filter & exp. time & field & pixel size &
seeing \\
          &            &        & (s)       & arcmin$^2$ & arcsec &
arcsec \\
          &            &        &           &       & & \\
\hline
          &            &        &           &       & & \\
CFHT & Megacam & $g'$  & 4830 & 60$\times$60   & 0.186 & 0.9 \\
     &         & $r'$  & 2450 & 60$\times$60   & 0.186 & 0.9 \\
VLT  & FORS2   & $z'$  &   90 & 9.9$\times$9.1 & 0.252 & 0.64 \\
SOAR & SOI     & R     &  480 & 5$\times$5     & 0.154 & 1.0 \\
     &         & Gal.H$\alpha$ & 360  & 5$\times$5 & 0.154 & 1.0 \\
     &         & Gal.[SII]     & 2460 & 5$\times$5 & 0.154 & 1.0 \\
          &            &        &           &       & & \\
\hline
\hline
\end{tabular}
\end{center}
\label{tab:obs}
\end{table*}

We obtained optical images in various bands with several telescopes,
as summarized in Table~\ref{tab:obs}. 

At the \oph\ cluster redshift, the H$\alpha$ line falls in the
wavelength range covered by the filter adapted to observe the Galactic
[SII]6717,6731 emission, so we observed \oph\ in this filter with the
SOAR telescope and SOI instrument during an observing run dedicated to
galaxy clusters in 2009. We also took brief exposures in the
Galactic H$\alpha$ and R band filters, to allow a good continuum
subtraction in order to derive the H$\alpha$ emission and the star
formation rate in the galaxies belonging to the cluster.

\subsection{Galaxy selection}

    As a pilot survey of galaxies, we chose an area of $10.1\times
9.3$~arcmin$^2$ in RA and DEC centered on the cD galaxy.  This area
covers the whole VLT and SOAR fields.

At the beginning, we tried an automated galaxy survey, but failed,
because too many blended stars were selected erroneously. So instead, 
we conducted our galaxy survey by eye inspection on the combined
$g'$ and $r'$ band images by changing the image brightness and
contrast on the ds9 and Gaia viewers. The detections of galaxies of
small angular sizes are severely disturbed by many foreground stars.
To overcome this difficulty, we made an eye inspection not only of the
original images but also of the star-subtracted images (see section
2.3).

On these star-subtracted images, extended objects of small angular
sizes show donut-shaped structures around their removed cores, while
stars are clearly removed.  This is our discrimination between stars
and galaxies.  For objects difficult to classify, we refered to
stellarity indexes deduced from SExtractor and to results of image
profiles measured with ``imexamine" in IRAF. We selected 225
  objects, and later closely re-examined 63 of these objects. Out of
  those 63 objects, 2/3 are found to be residuals around bright stars
  (see in section 2.3) or objects with stellarity index $> 0.15$, and
  the remaining 1/3 are too faint to obtain reliable stellarity
  indexes.  In Table A.1, we list 162 objects likely to be galaxies
  with their coordinates and magnitudes.  Although the completeness of
  our survey is difficult to estimate in this complex area, we roughly
  estimate it to be higher than 70\% for galaxies with angular sizes
  larger than 2.5 to 3~arcsec. Indeed, we independently made a galaxy
  selection on the VLT image which has the best seeing, and found only
  two additional galaxies.  Some small galaxies lying close to bright
  stars are missed, because they are buried in diffraction spiders and
  reflexion residuals of bright stars. Besides, some dwarf galaxies of
  low surface brightnesses having diameters larger than 3~arcsec are
  also missed, because they are buried in irregularities of the sky
  background.

\subsection{The method to subtract stars from the images}

\begin{figure*} 
\centering \mbox{\psfig{figure=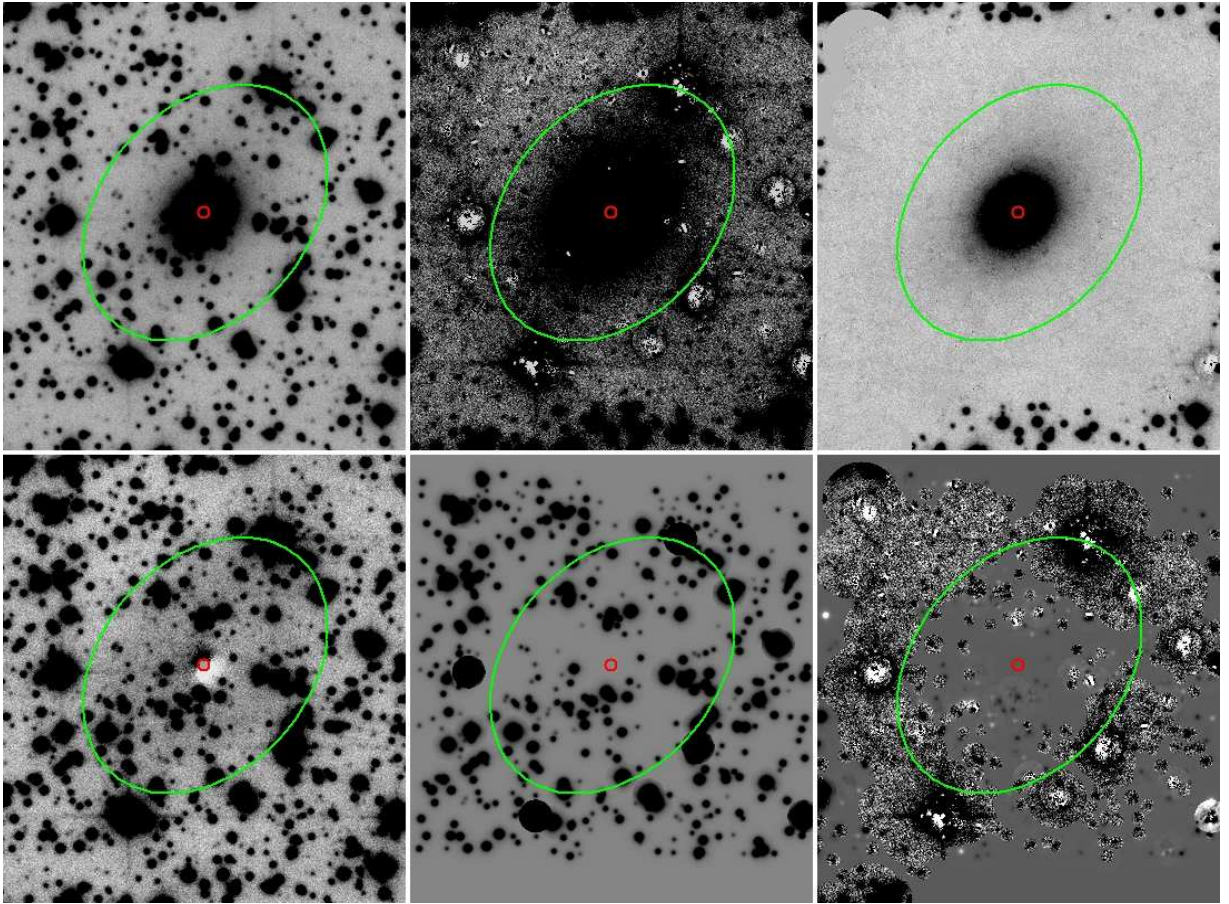,width=16cm}}
\caption{ Images illustrating our method for star-subtraction. The
  field of view of each image is about $1.1\times 1.3$~arcmin$^2$ in
  the east-west and north-south directions, respectively.  The small
  red circles indicate the position of the nucleus and the green
  ellipses represent the outer boundary of the galaxy.  (a: top
  left)~Original $r'$ band image of 2MASX J17121895-2322192 =
  OPH~171218.98-232219.1, one of the brightest galaxies.  (b: bottom
  left) Stellar component after subtracting the galaxy model from
  image (a). This image is almost free of galaxy light, and so
  appropriate for evaluation of stellar parameters and measuring the
  PSFs.  (c: top middle) Extracted galaxy image after automated
  star-subtraction with PSFs.  Since irregular residuals of saturated
  stars are not well removed, they disturb the galaxy profile and its
  surrounding sky area.  (d: bottom middle) Image showing the
  automatically subtracted stars.  This is the difference between
  images (a) and (c) and shows that more than 100 stars superposed on
  the galaxy are removed.  (e: top right) Final image of the galaxy
  after manual cleaning of residuals around saturated stars in image
  (c).  (f: bottom right) Residuals left after manual cleaning.  This
  is the difference between images (c) and (e).  }
\label{fig:star_sub}
\end{figure*}

On our $r'$ band Megacam image, the number density of foreground stars
which disturb the galaxy surface photometry in Ophiuchus is found to
be about 350 stars~arcmin$^{-2}$, and therefore 100-200 stars are
superposed on large bright galaxies.  Hence, star-subtraction is
essential to achieve good quality galaxy photometry in Ophiuchus.

    The present star subtraction procedure is a modified version of
the method originally developed by Nagayama et al. (2006). It consists
of: 1)~extraction of the point spread function (PSF),
2)~star subtraction, and 3)~cleaning of residuals such as those due to
saturated stars or to diffraction on the spider.  All these procedures
were done using the ``DAOPHOT" and ``STSDAS" packages in IRAF. 
 
We first describe the extraction of the PSF. Since almost all the
stars that should be removed are blended with one another due to the
high star density in the Galactic disk, a precise determination of the
PSF is crucial.  This requires isolated non-blended stars.  However,
almost all PSF candidate stars within an appropriate magnitude range
are blended.  So we obtained the final PSF after iterating three times
the deblending process for these stars.  We detected a spatial
variation of the PSF even within a few arcminutes, due not only to the
optics of the CFHT wide-field Megacam camera, but also to the
different performances of the mosaic CCDs of the camera.  Therefore,
star subtraction was processed by considering locally extracted PSFs.

Simple star subtraction from an original image is not appropriate,
because the profiles of stars superimposed on a galaxy are disturbed
by the light of the target galaxy.  So, before star subtraction, the
galaxy light should be removed from the original image by subtracting
a brightness model of the galaxy.  This model is extracted from a
crude surface photometry of the galaxy with the ``ellipse" and
``bmodel" tasks in the IRAF STSDAS package.  After subtracting this
model galaxy from the original image, an image without the galaxy
component is created (Fig.~2b). Based on this image, the stellar
parameters for star subtraction with the series of PSFs are calculated
with the ``daofind", ``phot", etc. tasks.  By subtracting these stars
to the original sub-image with the ``allstar" task, we then obtain a
galaxy image without the superposition of foreground stars.  Starting
with this crudely removed galaxy image, we repeat this loop two or
three times (Fig.~2c).

     The automated star-subtraction described above is almost perfect
for non-saturated stars superposed on galaxies even if they are
blended in a complicated way (Fig.~2c).  However, this process works
very poorly for saturated stars, which are always
accompanied by large residuals, e.g., scattered light, spider
diffraction patterns, etc.  These have irregular shapes, so we are
obliged to remove them manually one by one with the ``imedit" task in
IRAF.  If the outer annular zone encircling these residuals is clean
and not disturbed by other residuals, the aperture around the
residuals is replaced by an aperture interpolated from this boundary
annulus with replacement option ``b" in ``imedit".  If replacement by
interpolation is not appropriate due to neighboring residuals or
diffraction patterns, the residual aperture is substituted, with the
replacement option ``m", by a nearby aperture region of similar surface
brightness to that of the galaxies (Fig.~2e).  This cleaning process
is the most delicate and painstaking job, since it must not modify the
luminosity profiles and total magnitudes of the galaxies.

     The error sources on the star subtraction processes for galaxy
photometry are: 1)~the subtraction of non-saturated stars with PSFs,
2)~the replacement of residuals by interpolation and/or substitution as
described above.  Besides, bright stars located around the peripheries
of galaxies cause uncertainty on the integration boundary for galaxy
photometry.  For elliptical and S0 galaxies which have smooth and
symmetric structures, interpolation and substitution may not be a
serious problem, but for spiral galaxies with knotty arms and for
interacting galaxies with asymmetric tails and bridges, the situation
is more difficult.  Fortunately, the Ophiuchus cluster is of cD-type
and therefore rich in E and S0 galaxies, and poor in spiral and
interacting galaxies.  To estimate errors on our photometric
measurements, we considered two sets of images with different
star-subtractions, and made photometric measurements twice for each
band.  By comparing these two results, we estimate the errors 
in the three measured bands as follows:
\begin{itemize}
\item  0.07, 0.14, and 0.25~mag for objects with $r' < 19$, $19 < r' < 21$, 
and  $r'>21$, respectively, 
\item  0.08,  0.15, and 0.25 mag for objects with $g' < 20$, $20 < g' < 22$, 
and  $g'>22$ respectively, and
\item  0.05,  0.10, and 0.25 mag for objects with $z' < 20$, $20 < z' < 21$, 
and  $z'>21$ respectively.
\end{itemize}

These processes are illustrated in Fig.~\ref{fig:star_sub}.

\subsection{The photometric catalogue}

After subtracting the star contribution (as described above), we
extracted small subimages containing each galaxy and its immediate
surroundings, and measured magnitudes and morphological parameters
with SExtractor (Bertin \& Arnouts 1996). This was done in the $g'$,
$r'$, and $z'$ bands for all the galaxies located in the field covered
by the VLT/FORS2 image. Good quality measurements were achieved for
162 galaxies in the $g'$, $r'$, and $z'$ bands respectively. In the
region covered by Megacam but not by FORS2, only $r'$ band magnitudes
were measured for the galaxies with spectroscopic redshifts, in order
to have a sample as large as possible of galaxies with both
spectroscopic redshifts and magnitudes, to apply the Serna \& Gerbal
analysis (see Section~\ref{sec:SG}).  The photometric catalogues are
given in Tables~\ref{tab:photo_small} and \ref{tab:photo_r_ext}.

\subsection{The spectroscopic catalogue}

Redshifts of bright galaxies were obtained covering a very large area
on the sky with the AAO 6dF, CTIO 1.5m and Lick 3m telescopes (Wakamatsu
et al. 2005), and are listed in Table~\ref{tab:redshifts}.

We also obtained spectroscopic data with the ESO VLT/UT1 telescope
using FORS2 in two adjacent fields covering the very central region of
the cluster, for a total exposure time of 1500s.  Spectra were
obtained for 55 objects, out of which 54 gave reliable redshifts.  Out
of these, 17 proved to be galaxy spectra (about 14 belonging to the
\oph\ cluster) and 37 were stars.

\begin{figure} 
\centering \mbox{\psfig{figure=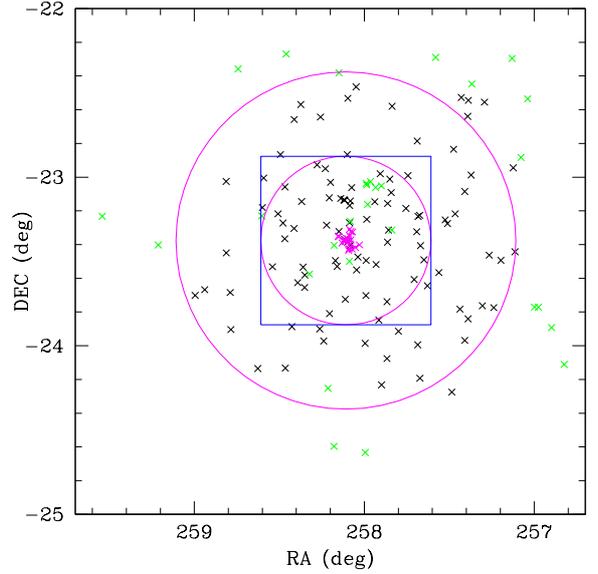,width=8cm}}
\caption{Spatial distribution of the galaxies with spectroscopic
  redshifts, color-coded as follows: magenta crosses: our FORS2
  measurements, green crosses: Hasegawa measurements, black crosses:
  values found in NED. The two magenta circles have radii of 0.5~deg
  (1~Mpc) and 1~deg (2~Mpc) and the blue square shows the size of the
  Megacam $g'$ and $r'$ band images.}
\label{fig:xyzspec}
\end{figure}

We retrieved redshifts in the NED data base in a region 
shown in Fig.~\ref{fig:xyzspec}.

We also retrieved  a spectrum of the cD galaxy taken by
John Huchra at CTIO and measured its velocity to be cz=$8844\pm
60$~km/s.  This is within the error of the systemic velocity of the
cluster $8878 \pm 76$ km/s (as inferred from the dynamical analysis,
see Sect.~4.2).

\begin{figure} 
\centering \mbox{\psfig{figure=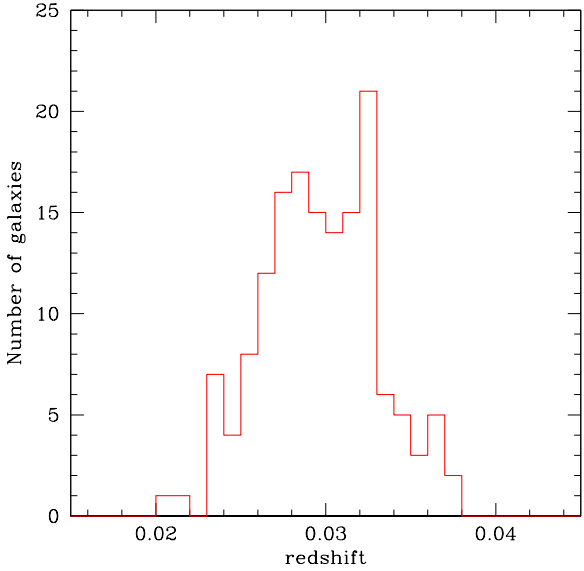,width=8cm}}
\caption{Spectroscopic redshift histogram for the 152 galaxies of
  Table~A3, with measured redshifts in the $0.02\leq z \leq 0.04$
  redshift range.}
\label{fig:histoz}
\end{figure}

We built a spectroscopic redshift catalogue by combining the Wakamatsu
et al. (2005) measurements and our FORS2 measurements, and adding
redshifts available in NED for galaxies which were not in the two
previous catalogues. The resulting catalogue was limited to the
redshift range $0.02 \leq z \leq 0.04$, somewhat broader than the
cluster range in order not to ``lose'' galaxies. For the galaxies with
several redshift measurements, we took in decreasing priority the
FORS2 value, the Wakamatsu et al. (2005) value, and the NED value. Our
final redshift catalogue is given in Table~\ref{tab:redshifts} and
includes 152 galaxies after eliminating objects with multiple
measurements.  We estimate the uncertainties on the corresponding cz
velocities to be smaller than 280~km~s$^{-1}$ for the FORS2 data (see
Adami et al. 2011) and between 100 and 300~km~s$^{-1}$ for the
Wakamatsu et al. (2005) data.  The spatial distribution of the
galaxies with measured redshifts is displayed in
Fig.~\ref{fig:xyzspec}, and the redshift histogram is shown in
Fig.~\ref{fig:histoz}.

For the galaxies with spectroscopic redshifts located within the
CFHT/Megacam field of view but outside the FORS2 field, we performed
the star subtraction and measured the magnitudes in the $r'$ band as
described above. This gave a catalogue of 89 galaxies with both
redshifts and $r'$ band magnitudes within the $1\times 1$~deg$^2$
Megacam area, to which we applied the Serna \& Gerbal software to
search for substructures (see section~\ref{sec:SG}).

\section{Results}

\subsection{The red sequence}

\begin{figure*} 
\centering \mbox{\psfig{figure=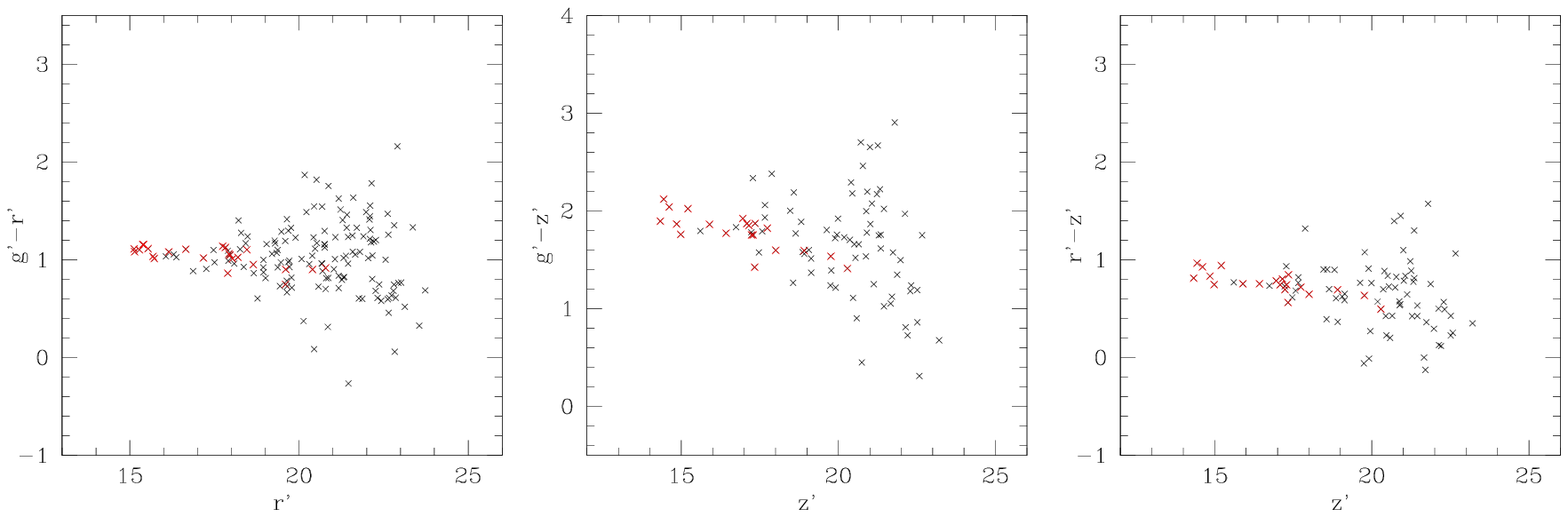,width=16cm,clip=true}}
\caption{Colour-magnitude diagrams. The red crosses are the galaxies with
spectroscopic redshifts in the $0.02 \leq z \leq 0.04$ range.}
\label{fig:colmag}
\end{figure*}

For the galaxies with measured $g'$, $r'$ and $z'$ magnitudes, we drew
the three possible colour-magnitude diagrams. These are shown in
Fig.~\ref{fig:colmag}. We can see from the positions of the galaxies
with spectroscopic redshifts in these diagrams that the red sequence
is very well defined in all three plots. This confirms the validity of
our photometric treatment and star subtraction. 

The best fit to the $(g'-r')$ versus $r'$ relation computed with the galaxies
brighter than $r'=20$ is :
$$ (g'-r')\ =\ (-0.03955\pm 0.0114)r'\ +\ (1.72\pm 0.21).$$
This relation can be compared to that estimated for a cluster of
comparable richness and redshift, such as Coma. Eisenhardt et
al. (2007) give the following relation for Coma (see their Table~13):
$$ (B-R)\ = (-0.055\pm 0.006)R\ +\ (2.259\pm 0.074).$$
If we translate our relation to B and R magnitudes using the Fukugita et al. (1995)
transformations, we find for \oph : 
$$ (B-R)\ = (-0.03955\pm 0.0114)R\ +\ (2.54\pm 0.21). $$
Though the slope for \oph\ is flatter than for Coma, the slopes are
consistent within error bars, and the normalisation parameters are of
the same order of magnitude.  

We can also note from the colour-magnitude relations that there are
very few bright blue galaxies in \oph : in the $(g'-r')$ versus $r'$
diagram, for example, there is not a single galaxy below the red
sequence for $r'\leq 19.5$.  This result is comparable to that found,
for example, in the Coma cluster, where few galaxies bluer than the
red sequence are detected (see Fig.~17 in Adami et al. 2007).
 
This agrees with our idea that Ophiuchus is an ``old'' cluster, and
this is also confirmed by the very small number of galaxies showing
ongoing star formation (see next subsection).

\subsection{Star formation in Ophiuchus galaxies}

\begin{figure}[!ht]
  \begin{center}
    \includegraphics[angle=270,width=3.5in]{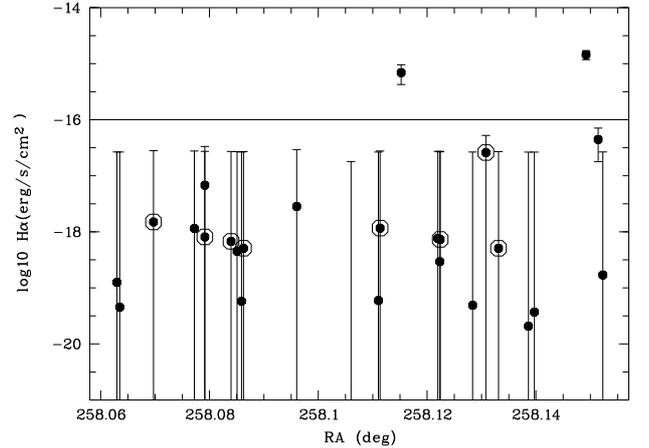}
    \caption{H$\alpha$ calibrated fluxes (in units of
      erg~s$^{-1}$~cm$^{-2}$, in logarithmic scale) versus RA
      coordinates for all the galaxies detected along the Ophiuchus
      line of sight. Large open circles are the galaxies observed
      spectroscopically. The horizontal line shows the detection limit
      at $10^{-16}$~erg~s$^{-1}$~cm$^{-2}$.}
  \label{fig:alphahalpha}
  \end{center}
\end{figure}

As explained in the introduction, we expect the galaxies of \oph\ to
be in majority old, and therefore lacking star formation. To test this
hypothesis, we observed the cluster in a narrow band filter containing
H$\alpha$ and [NII] at the cluster redshift, as explained in
Section~\ref{sec:obs}.  We consider in the following H$\alpha$ and
[NII] as a single line noted H$\alpha$.

We first renormalize the H$\alpha$, [SII], and R band SOAR fluxes to
take into account the different passbands of these filters and the
different exposure times. 

For a given object in Ophiuchus, we note I the flux measured in the
H$\alpha$ filter. It includes the Milky Way H$\alpha$ flux
(MWH$\alpha$) plus the continuum of the observed object at the filter
wavelength (C1).  We note II the flux received by the [SII] filter. It
includes the Milky Way [SII] flux (MWSII), plus the H$\alpha$ emission
of the observed object (OH$\alpha$), plus the continuum of the observed
object at the filter wavelength (C2).

\begin{equation}
{\rm I = C1 + MWH}\alpha
\end{equation}

\begin{equation}
{\rm II = C2 + OH}\alpha + {\rm MWSII} 
\end{equation}

Our goal is to measure the H$\alpha$ emission of the observed object:
OH$\alpha$.

From Fig.~8 of Haffner et al. (1999), we estimate the ratio between
the Milky Way H$\alpha$ and [SII] emissions to be 4.7$\pm$0.3 at the
Ophiuchus position:
\begin{equation}
{\rm MWH}\alpha / {\rm MWSII} = 4.7\pm 0.3 
\end{equation}

Moreover, we can safely assume that C1$\simeq$C2, since the wavelengths
of the H$\alpha$ and [SII] filters are very similar:
\begin{equation}
{\rm C1 = C2 = C}
\end{equation}

Combining equations (1) to (4), we have:
\begin{equation}
{\rm OH}\alpha = {\rm II - (I/4.7) - C*(1.-(1./4.7))}
\end{equation}

We must now estimate C (our H$\alpha$ and [SII] images and our spectra
are not flux calibrated).  We assume this value to be the same for all
the Ophiuchus galaxies, a reasonable hypothesis since we are
considering a very homogeneous population of cluster galaxies. We then
select the galaxies for which no H$\alpha$ line is visible in our
spectra. In this case, equation (5) becomes:
\begin{equation}
{\rm C = (II - (I/4.7)) /(1.-(1./4.7))}
\end{equation}

We then estimate C for the previously selected Ophiuchus galaxies with
no H$\alpha$ emission, and apply this value to all the considered
objects. We therefore have a direct access to the uncalibrated
OH$\alpha$ flux with equation (5) for all the observed objects.

To have a calibrated flux, we compare our SOAR R-band and CFHTLS r'
fluxes and apply the corresponding zero point to OH$\alpha$
fluxes. We then generate Fig.~\ref{fig:alphahalpha}.  Considering the
galaxies for which we have a spectrum (we did not detect any H$\alpha$
lines in our spectroscopic sample), we estimate the maximum value
below which the H$\alpha$ emission is not detectable to be about
$10^{-16}$~erg~s$^{-1}$~cm$^{-2}$.

Only two galaxies in Fig.~\ref{fig:alphahalpha} show a significant
H$\alpha$ emission. Their coordinates are (258.149$^\circ$,
$-23.3223^\circ$) and (258.115$^\circ$, $-23.3696^\circ$) and their
respective H$\alpha$ fluxes are $1.446\times 10^{-15}$ and
$0.692\times 10^{-15}$~erg~s$^{-1}$~cm$^{-2}$.

After converting H$\alpha$ fluxes into star formation rates (hereafter
SFR) with the Kennicutt (1998) relation, our detection limit
translates to $1.5\times 10^{-3}$~M$_\odot$~yr$^{-1}$, a fairly low
value.  Even the two galaxies in which we do detect H$\alpha$ have
fairly low SFRs of the order of  0.02~M$_\odot$~yr$^{-1}$
and 0.01~M$_\odot$~yr$^{-1}$ respectively.

We therefore conclude that the Ophiuchus galaxies have very low SFRs,
as expected.

\subsection{The cluster internal structure and dynamics}

\subsubsection{Selection of spectroscopic cluster members}

\begin{figure}[!ht]
  \begin{center}
    \includegraphics[width=3.5in]{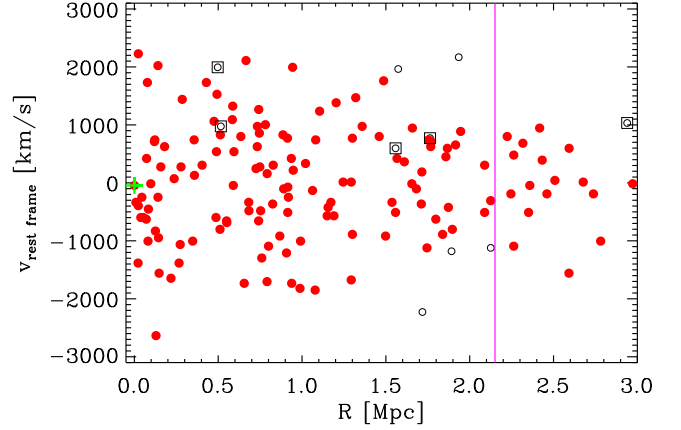}
    \caption{Rest-frame velocities vs. distances from the cluster
      center for all galaxies with redshifts in the cluster region.
      Filled (red) dots identify cluster members; the green-crossed
      red dot is the cD galaxy. The empty circles identify the 5 interlopers
      found by the shifting gapper procedure. The empty circles
      within squares identify the 5 interlopers found by the
      adaptive-kernel procedure. Note that some of them are nearly
      coincident in projected phase-space, hence the symbols are
      superposed.}
  \label{fig:rv}
  \end{center}
\end{figure}

We examined the positions of the galaxies in the ``blue'' peak
(i.e. those with $0.02 \leq z < 0.032$) and those in the ``red'' peak
($0.032 \leq z \leq 0.04$) of Fig.~\ref{fig:histoz}, but were not able
to separate them spatially. Therefore, if the blue and red
distributions correspond to two clusters in the process of merging,
the merger must be taking place along a direction close to
perpendicular to the plane of the sky.

To identify the cluster members among the 152 galaxies with measured
redshift, we use the 'shifting gapper' method of Fadda et al. (1996).
This method searches for gaps of 1000 km~s$^{-1}$ in the velocity
distributions, within overlapping radial intervals of 0.4 h$^{-1}$ Mpc
(corresponding to 0.56 Mpc for our adopted cosmology). These gaps are
used to separate interlopers from clusters members. Five galaxies are
identified as interlopers by this procedure.

We then refine our membership identification by another procedure
which is similar to the one developed by Biviano et al. (1996) for the
Coma cluster. We determine two adaptive kernel maps of the galaxy
number densities, one unweighted, and another weighted by the galaxy
rest-frame velocities. The ratio of the two maps defines an adaptive
kernel map of local velocities.  We then run 1000 bootstrap
resamplings to establish the significance of the adaptive kernel map
of galaxy velocities, at each galaxy position, $\sigma_{boot,i}$ (see
Appendix A of Biviano et al. 1996 for more details). When the adaptive
kernel value of the velocity at the $i$-th galaxy position deviates
from zero at more than $3 \sigma_{boot,i}$, we reject galaxy $i$ from
the list of cluster members.  We reject five galaxies by this
  procedure, so we are left with a total of 142 cluster members.

The projected-phase space distribution of all galaxies with
spectroscopic redshifts in the cluster region is shown in
Fig.~\ref{fig:rv}. It displays a decrease of velocity dispersion with
increasing distance from the cluster center, as seen in many nearby
clusters (e.g. Biviano et al.  1997). 

\subsubsection{Dynamical analysis}

\begin{figure}[!ht]
  \begin{center}
    \includegraphics[width=3.5in]{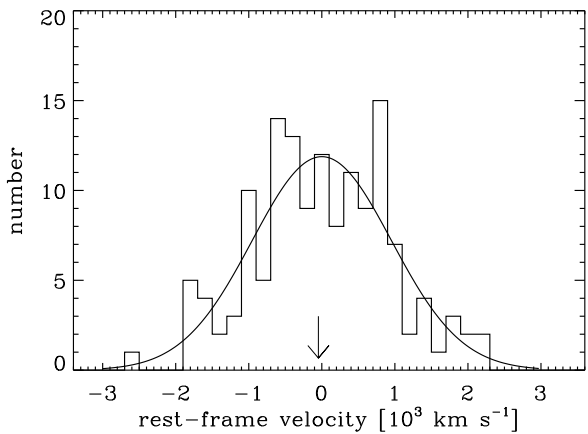}
    \caption{Histogram of the rest-frame velocities of the 
      identified cluster members. The best-fitting gaussian is
      overplotted. The arrow indicates the velocity of the cD galaxy.}
  \label{fig:hv}
  \end{center}
\end{figure}

The histogram of rest-frame velocities for the
cluster members is
shown in Fig.~\ref{fig:hv}. The mean redshift and velocity dispersion
of the cluster members are $\overline{z}=0.0296 \pm 0.0003$ and
$\sigma_{\rm{v}}=954_{-55}^{+58}$ km~s$^{-1}$, respectively (derived
using the robust biweight estimator, see Beers et al. 1990).  Our new
estimate of $\sigma_v$ is significantly lower than the value obtained
by Wakamatsu et al. ($1050 \pm 50$ km~s$^{-1}$).  Using our new values
of $\overline{z}$ and $\sigma_v$ we draw the corresponding Gaussian
distribution in Fig.~\ref{fig:hv}. A Kolmogorov-Smirnov test
(e.g. Press et al. 1992) gives a probability of 0.82 that the observed
velocity distribution is a random draw of the Gaussian
distribution. In other terms, the observed distribution is
statistically indistinguishable from a Gaussian. This is generally
taken as evidence for a relaxed dynamical configuration (e.g. Girardi
et al. 1993; Ribeiro et al. 2011).

Additional support for a dynamically relaxed status of the
cluster also comes from the position of the cD in projected
phase-space (Beers et al. 1991). It is in fact spatially coincident
with the peak of the X-ray emission, and its velocity is consistent
within the errors with the mean cluster velocity, $\Delta \rm{v}=47 \pm 97$
km~s$^{-1}$.

It is possible to obtain a first, preliminary estimate of the cluster
mass from the $\sigma_{\rm{v}}$ estimate, via the scaling relation of
Munari et al. (2013; eq. 1): $M_{200}=9.3 \, [8.1,10.7] \times 10^{14}
\, M_{\odot}$, corresponding to $r_{200}=2.0 \, [1.89,2.07]$~Mpc,
where the values in brackets correspond to the 1 $\sigma$ confidence
intervals \footnote{The virial radius $r_{200}$ is the radius of a
  sphere with mass overdensity 200 times the critical density of the
  Universe at the cluster redshift. The virial mass $M_{200}$ is
  directly related to $r_{200}$ via $M_{200} \equiv 100 \, H_z^2 \,
  r_{200}^3/G$, where $H_z$ is the Hubble constant at the cluster mean
  redshift.}.

We proceed by estimating the cluster mass profile using the MAMPOSSt
technique of Mamon et al. (2013). This technique performs a Maximum
Likelihood fit of selected mass and velocity-anisotropy models, to the
projected phase-space distribution of cluster members. 

We consider three models for the mass distribution: 1) Burkert (1995;
'Bur'), 2) Hernquist (1990; 'Her'), and 3) the popular Navarro et
al. (1996; 'NFW'). The three models are characterized by different
logarithmic slopes of their mass density profiles, 0, $-1$, and $-1$
near the center, and $-3, -4, -3$ at large radii, for Bur, Her, and
NFW, respectively. Each of these models is characterized by two free
parameters, the virial radius $r_{200}$, and the scale radius.  The NFW
scale radius corresponds to the radius where the logarithmic slope of
the mass density profile equals $-2$, and we therefore denote it by
$r_{-2}$. The Bur scale radius approximately corresponds to $2/3 \,
r_{-2}$, while the Her scale radius corresponds exactly to $2 \,
r_{-2}$. For the sake of homogeneity, we rescale the scale radii of
the Bur and Her model to always quote the results for $r_{-2}$.

We consider two models for the velocity anisotropy profile, 
\begin{equation}
\beta(r) = 1 - {\sigma_\theta^2(r) + \sigma_\phi^2(r)  \over
  2\,\sigma_r^2(r)} = 1 - {\sigma_\theta^2(r) \over \sigma_r^2(r)}  
\label{e:beta}
\end{equation}
where $\sigma_\theta, \sigma_\phi$ are the two tangential components,
and $\sigma_r$ the radial component, of the velocity dispersion, and
the last equivalence is obtained in the case of spherical symmetry.
Negative, null, and positive values of $\beta$ correspond to galaxy
orbits that are tangential, isotropic, and radial, respectively.  One
of the two models ('C') assumes constant $\beta(r)$ at all radii. The
other ('T' from Tiret et al. 2007) is of the form: $\beta(r)=\beta_{\infty}
\, r/(r+r_{-2})$.  The two models are both characterized by only one
free parameter, the constant value of $\beta$ for the C model, and
$\beta_{\infty}$ for the T model, since $r_{-2}$ in this model is the
same scale radius parameter of the mass models.

In MAMPOSSt, and in general in all dynamical analyses based on the
Jeans equation (Binney \& Tremaine 1987), knowledge of the radial
dependence of the completeness of the spectroscopic sample is
required. An unknown or uncorrected-for radial-dependent
incompleteness would bias the determination of the number density
profile of the tracer of the gravitational potential, which enters the
Jeans equation. On the other hand, the velocity distribution of the
tracers is in general unaffected by incompleteness problems, since
observations do not generally select cluster members in velocity
space. In the case of Ophiuchus, an estimate of the radial
completeness of the spectroscopic sample is not available, because of
the stellar crowded field. As a consequence we must make the
assumption that the number density profile of the tracers and the mass
density profile of the cluster have the same shape, and are therefore
characterized by the same scale parameter, $r_{-2}$. This case was
already considered by Mamon et al. (2013), and denoted 'TLM' (for
'Tied Light and Mass'). They showed that MAMPOSSt can work very well
also within this restrictive assumption.

\begin{table}
\caption{Results of the MAMPOSSt analysis}
\label{tab:mamposst}
\begin{tabular}{lc}
\hline 
\hline
 & \\
$r_{200}$ [Mpc] & $2.1 \, [1.8, 2.2]$ \\
$r_{-2}$ [Mpc] & $0.7 \, [0.5,1.4]$ \\
$\beta_{\infty}$ & $0.8 \, [0.4,1.0]$ \\
$M_{200}$ [$10^{14} \, M_{\odot}$] & $11.1 \, (6.8,13.7)$ \\
 & \\
\hline
\end{tabular}
\tablefoot{Results are shown for the best-fit models Her+T. Values in
brackets are 1 $\sigma$ lower and upper confidence levels.}
\end{table}

We find that Her+T provides the best-fit among the six combinations of
the three mass and the two anisotropy models. However, all models are
statistically acceptable. The best-fit values for the three free
parameters $r_{200}, r_{-2}, \beta_{\infty}$ are listed in
Table~\ref{tab:mamposst}, with their uncertainties, obtained by
marginalizing each parameter with respect to the other two. For
completeness we also translate the constraints on $r_{200}$ into
constraints on the cluster mass, $M_{200}$. The values of
$\beta_{\infty}$ indicate that the galaxy orbits are mostly radially
elongated in this cluster.

\begin{figure}[!ht]
  \begin{center}
    \includegraphics[width=3.5in]{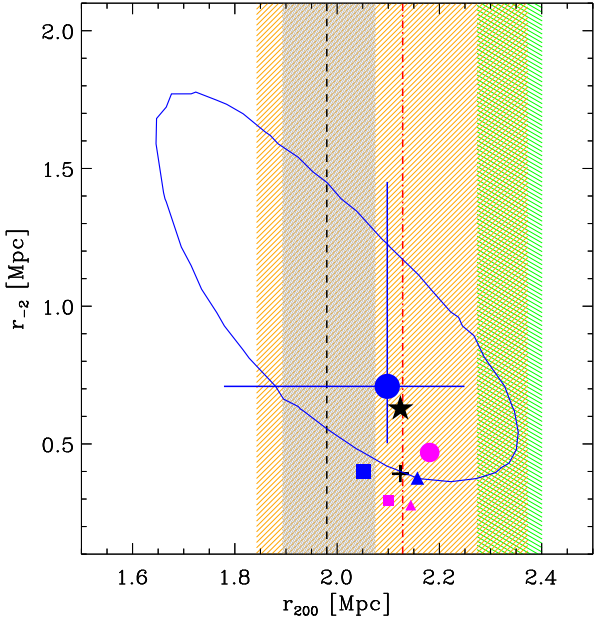}
    \caption{Results of the dynamical analysis in the $r_{-2}$ vs.
      $r_{200}$ plane. The squares, dots, and triangles indicate the
      best-fit values obtained by MAMPOSSt for the Her, NFW, and Bur
      mass models. Magenta and blue symbols are for the C and T $\beta(r)$
      models, respectively. The symbol size is proportional to the
      MAMPOSSt likelihood.  The ellipse indicates the 1 $\sigma$
      uncertainty on the best-fit MAMPOSSt results, after
      marginalization over the velocity anisotropy parameter. The
      error bars are obtained upon marginalization of each parameter
      over the other two.  The cross indicates the mean of the
      values obtained by the different models considered for the
      MAMPOSSt analysis.  The vertical black dashed line indicates the
      value of $r_{200}$ obtained from the cluster velocity dispersion
      using the scaling relation of Munari et al. (2013). The shaded
      grey region indicates the uncertainties on this value.  The
      vertical red dash-dotted line indicates the value of $r_{200}$
      obtained with the Caustic technique of Diaferio \& Geller
      (1997). The shaded orange region indicates the uncertainties on
      this value. The shaded green region indicates the range of
      $r_{200}$ as obtained from the range of X-ray temperatures found
      in the literature (see text), using
      the scaling relation of Vikhlinin et al. (2009).
      The star symbol represents the theoretical expectation from De
      Boni et al. (2013), obtained adopting the average value of $r_{200}$.}
  \label{fig:rvrs}
  \end{center}
\end{figure}

In Fig.~\ref{fig:rvrs} we show the constraints obtained by MAMPOSSt in
the $r_{-2}$ vs. $r_{200}$ plane. The uncertainties on the $r_{200}$
value of the best-fit models, Her+T, are larger than the differences
among the values obtained with the different models.  On the other
hand, some of the $r_{-2}$ values are more than 2 $\sigma$ below the
value obtained with the Her+T model. The average $[r_{200},r_{-2}]$
value (the '+' symbol in the figure) among the different models is in
fact below the value of the best-fit model. However, it is remarkable
that the latter is in excellent agreement with the theoretically
predicted value obtained using the concentration-mass relation of De
Boni et al. (2013, the star symbol in the figure), and the average
value of $r_{200}$, because the theoretical estimate comes from a
concentration-mass relation. Given $r_{200}$, one has the mass, and from
the mass, the concentration. Given the concentration and $r_{200}$, one has
 $r_{-2}$. 

We also display in Fig.~\ref{fig:rvrs} the $r_{200}$ values obtained
from $\sigma_{\rm{v}}$ using the scaling relation of Munari et
al. (2013). The relatively narrow confidence region (gray shaded area
in the figure) only reflects the uncertainty in the theoretical value
of the scaling relation. In reality, this is an underestimate, because
several systematic uncertainties are not taken into account (see
Munari et al. 2013 for more details).  Considering that the formal
uncertainties on the $\sigma_{\rm{v}}$-determined $r_{200}$ values are
smaller than the real ones, these values can be considered to be in
fair agreement with the MAMPOSSt results.

The MAMPOSSt results are also in agreement with the $r_{200}$ values
obtained by applying the Caustic technique of Diaferio \& Geller
(1997; see also Diaferio 1999) to our spectroscopic data-set (red line
and orange shaded region in Fig.~\ref{fig:rvrs}). This technique does
not require the assumption of dynamical equilibrium, but suffers from
an uncertain calibration which is generally determined using numerical
simulations. We adopt here the caustic mass calibration factor of Gifford
et al. (2013), ${\cal F}_{\beta}=0.65$.

\begin{figure}[!ht]
  \begin{center}
    \includegraphics[width=3.5in]{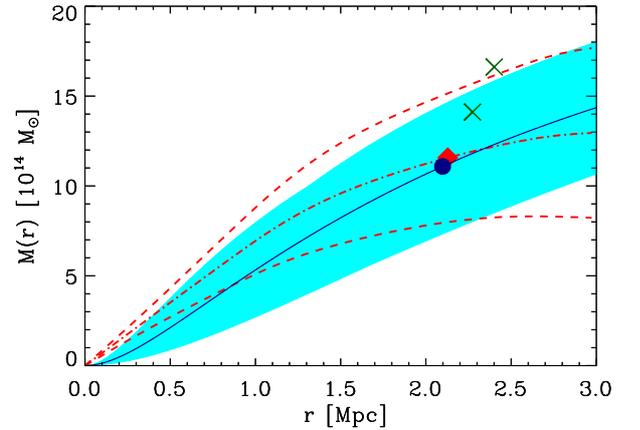}
    \caption{The cluster mass profile as obtained from MAMPOSSt (solid
      blue line) within its 1 $\sigma$ confidence interval (dashed
      cyan region). The blue dot is the location of
      $[r_{200},M_{200}]$.  The dash-dotted and dashed red lines
      indicate the cluster mass profile obtained from the Caustic
      technique and, its 1 $\sigma$ upper and lower limits
      respectively. The red diamond is the location of
      $[r_{200},M_{200}]$. The green 'X' symbols indicate the
      locations of $[r_{200},M_{200}]$ obtained using the observed
      range of cluster $T_X$ and the scaling relation of Vikhlinin et
      al. (2009).}
  \label{fig:mprof}
  \end{center}
\end{figure}

Fig.~\ref{fig:mprof} shows the two mass profiles obtained by MAMPOSSt and
the Caustic technique. They are in agreement within their uncertainties,
except in the central region, where the Caustic mass profile exceeds the
MAMPOSSt-derived one. This is due to a well known bias of the Caustic technique,
that tends to overestimate the mass profile in the central regions
(Serra et al. 2011).

Overall we conclude that there is a very good agreement in the mass and
mass profile estimates obtained with different techniques based on 
the projected phase-space galaxy distribution in the cluster. 

It is also possible to determine $r_{200}$ (and therefore also
$M_{200}$) from the cluster X-ray temperature.  For this we use the
scaling relation of Vikhlinin et al. (2009) with fixed exponent
$\alpha=1.5$, and the most recent and accurate determinations of the
cluster X-ray temperatures, $T_X=9.1$ keV (Nevalainen et al. 2009) and
$T_X=9.7$ keV (Fujita et al. 2008). The derived range of values
accounts for the uncertainties in the normalization of the scaling
relation of Vikhlinin et al. (2009).  As it can be seen from
Figs.~\ref{fig:rvrs} and \ref{fig:mprof} the allowed range of
$r_{200}$ (and $M_{200}$) values obtained from the X-ray temperatures,
overlaps with the range of values we infer from the kinematic
analysis.  The X-ray derived values tend however to be somewhat higher
than the values derived from the kinematic analysis.  Possibly, the
observed cluster $T_X$ is somewhat increased by a past collision
between a cluster and a subcluster.

\subsubsection{The presence of substructures}
\label{sec:SG}

We made a dynamical analysis with the Serna-Gerbal technique
(hereafter SG, 1996 release, Serna \& Gerbal 1996), a hierarchical
code (based on spectroscopic redshifts and optical magnitudes)
designed to detect substructures in the optical. The SG method has proved
to be quite powerful to show evidence for substructures in nearby
clusters (see Abell~496: Durret et al. 2000; Coma: Adami et al. 2005;
Abell~85: Bou\'e et al. 2008) as well as in more distant ones
(Guennou et al. 2014). We applied it here to the catalogue of 89
galaxies with both redshifts and $r'$ band magnitudes within the
$1\times 1$~deg$^2$ Megacam area.

Assuming a value of the mass to luminosity ratio (here taken to be
100), the SG method allows to estimate the masses of the structures
that it detects.  Although the absolute masses are not accurate (the
typical uncertainty is clearly larger than $10^{14}$ M$_{\odot}$), the
mass ratios of the various structures are well determined (Guennou
et al. 2014).  The SG method has also been extensively tested on
simulations by Guennou (2012), in particular concerning the effect of
undersampling on mass determinations.

One of the parameters that can be chosen is the minimum number n of
galaxies in a structure. We have applied the SG method with n=10 and
n=5 and find similar results in both cases.  The total mass of the
system is found to be ${\rm M_{tot}}=3.7\times 10^{14}$~M$_\odot$.
This estimate refers to the mass within the Megacam area, i.e. within
a radius of $\simeq 1$ Mpc.  We will divide the masses of the detected
substructures by this quantity to estimate the percentages of this
total mass included in each of the substructures.

\begin{figure}[!h]
  \begin{center}
    \includegraphics[width=3.0in]{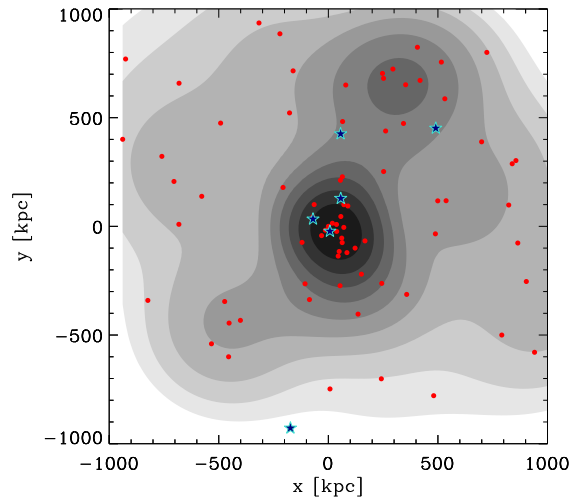}
    \caption{Adaptive-kernel number density map of the galaxies in
      structure 1 of the SG substructure analysis. The
      contours are logarithmically spaced. Red dots indicate the
      positions of the galaxies belonging to this group. Blue stars
      indicate the positions of the galaxies belonging to group 2 in
      the SG substructure analysis. }
  \label{fig:g1g2}
  \end{center}
\end{figure}

We find one large structure S1 of 75 galaxies and a much smaller
structure S2 of 13 galaxies. The mass ratios of these two structures
to the main cluster component are 0.67 and 0.065.  Structure S1 can
itself be divided into two substructures, S11 and S12, with 65 and 10
galaxies respectively, and S11 can itself be divided into two
structures.  In all cases we find a large substructure and a much
smaller one, implying that only minor mergers can be occuring.  These
results are illustrated in Fig.~\ref{fig:g1g2}, where an adaptive
kernel map of the main structure is drawn, with the galaxies of the
two structures S1 and S2 superimposed.

As an illustration of the relative significance of the structures
detected by the SG method, we refer the reader to Fig.~5 of Guennou et
al. (2014), showing the percentage of substructures detected by the SG
method as a function of the spectroscopic sampling.

The SG dynamical analysis therefore clearly confirms that \oph\ is not
a fully relaxed cluster, but any merger must have been minor, so it has
not strongly affected the dynamics of the cluster.

\section{Discussion and conclusions}

We now discuss the merging status of \oph. As mentioned above, only
Watanabe et al. (2001) claim that it is a major merger, based on a
temperature map drawn from ASCA data. On the other hand, several
studies, including the one presented here, strongly suggest that \oph\
is only the result of a minor merger.

From our data, there are several converging indices showing that \oph\
is not a cluster strongly perturbed by merging events.  First, the
center of the cD galaxy coincides with the X-ray peak. Second, the
velocity distribution is Gaussian, another evidence for global
relaxation, and third the cluster mass derived from the Jeans equation
agrees with the cluster mass computed with the caustic method, still
another evidence for relaxation. The masses derived from X-rays are a
little higher, but still consistent. Finally, the SG analysis does not
imply strong subclustering, given that a large fraction 
of the mass is not associated to any substructure.

\oph\ was originally believed to be a merging cluster, based on an
X-ray temperature map derived from ASCA data (Watanabe et al. 2001)
and showing two very hot regions to the West and South of the center,
some 20~arcmin away. The comparison of this map with simulations
suggested that a merger had occurred about 1~Gyr ago.  However, based
on Suzaku data, these results were contradicted by Fujita et
al. (2008), who did not detect the huge temperature variations found
by Watanabe et al. (2001). On the contrary, they found that \oph\ was
a cool core cluster, with isothermal gas (with a temperature
$kT=9.7_{-1.0}^{+0.9}$ keV) beyond 50~kpc from the cluster centre. We
can note however that even if \oph\ is a cool core cluster, it may
have experienced a minor merger, as indicated for example by the
simulations of Burns et al. (2008). Fujita et al. (2008) also wrote
that the iron-line ratios measured on the X-ray spectra indicated that
the ICM had reached an ionization equilibrium state, implying that the
cluster could not be a major merger.  However, this time based on
Chandra data, Million et al. (2010) found evidence for a collision
from the comet-like morphology of the X-ray emission near the cluster
centre. They interpreted the very strong temperature gradient, from
0.7~keV within 1~kpc to 10~keV at 30~kpc, as due to the fact that the
outer part of the cool core has been stripped by ram pressure.

Other results are consistent with \oph\ being a minor merger.  For
example, Murgia et al. (2009) have detected a mini radio-halo around
the cluster dominant radio galaxy, with a radio emissivity typical of
haloes in merging clusters. However, according to Zandanel et
al. (2014), the mini-halo in \oph\ has a radio luminosity much lower
than giant radio halos in merging clusters, so this suggests that any
merger can only have been a minor one.  The hard X-ray component
detected with XMM-Newton by Nevalainen et al. (2009) is indicative of
relativistic electrons, which are also responsible for the mini radio
halo.
The hypothesis of a minor merger agrees with the fact that Hamer et
al. (2012) measured an offset between optical line emission and the
BCG (about 2~kpc), and suggested it could be due to the merger-induced
motion of the ICM relative to the BCG. From their data, the merger
must have occured recently, about 20--100 Myr ago.

Adding this evidence together, it seems likely that a merger has
indeed occurred (probably quite recently), but that it was not a major
merger, and perhaps since Ophiuchus is so massive, this merger has not
perturbed much its dynamical state, though it may have affected the
intra-cluster medium to some exent. At smaller scale than the entire
cluster, some perturbations have been observed, such as an object
located 1.7~kpc from the BCG showing low ionization optical emission
lines and interpreted by Edwards et al. (2009) as a large cloud
falling towards the BCG. This system would then be comparable to that
described in the optical IFU observations of NGC~4696 by Farage et
al. (2010). However, they cannot perturb the cluster properties, as
for example confirmed by the lack of star forming galaxies.

\begin{acknowledgements}

We thank A. Boselli for discussions and the referee for suggestions.
FD acknowledges long-term support from CNES. 

\end{acknowledgements}

\appendix

\section{Catalogues}

The photometric catalogue in the $g'$, $r'$ and $z'$ bands of the
galaxies located within the VLT/FORS2 field is given in
Table~\ref{tab:photo_small} for 162 objects.

The photometric catalogue in the $r'$ band of the galaxies 
with a measured spectroscopic redshift but outside the VLT/FORS2 field
is given in Table~\ref{tab:photo_r_ext} for 65 galaxies.

The full catalogue of 152 spectroscopic redshifts (covering a region
larger than our images) is given in Table~\ref{tab:redshifts}.

\begin{table*} 
\centering
\setcounter{table}{0}
\caption{Photometric catalogue in the $r'$, $g'$ and $z'$ bands in the central
  region covered by the FORS2 $z'$ band image. The columns are: (1)~running number (the galaxies indicated with *
  have a spectroscopic redshift given in Table~\ref{tab:redshifts}), 
  (2)~object OPH name, (3) and (4)~RA and DEC (J2000.0), (5), (6) and (7)~$r'$, $g'$ and $z'$ band magnitudes (the 
  symbols : and :: respectively indicate a relatively large and a large uncertainty),
  (8)~quality of the galaxy classification (1:~almost definitely a galaxy, 2:~probably a galaxy, 3:~possibly a galaxy;
  when the object is certainly a galaxy, there is no indication in this column), 
  (9)~notes, other identifications and comments (poor photometric condition around object, e.g., bright stars 
  or diffraction on the spider, are expressed as 1, 2, and 3).
}
\begin{tabular}{rllllllll}
\hline
\hline
         &                           &             &             &           &          &          &         &                         \\
Running  & OPH                       &  RA         &  DEC        &    $r'$   &    $g'$  &    $z'$  & Quality & Note                    \\
number   & name                      &  (J2000.0)  &  (J2000.0)  &           &          &          &         &                         \\
         &                           &             &             &           &          &          &         &                         \\
\hline
         &                           &             &             &           &          &          &         &                         \\
     1~~ &  OPH J171205.08-232033  &  258.02116  &  -23.34253  &   16.872: &  17.756: &          &         &  1                      \\
     2~~ &  OPH J171205.41-232050  &  258.02255  &  -23.34746  &   20.756  &  22.050  &          &         &                         \\
     3~~ &  OPH J171205.68-232158  &  258.02367  &  -23.36624  &   21.588  &  22.669  &          &    1    &                         \\
     4~~ &  OPH J171205.85-232149  &  258.02439  &  -23.36387  &   17.261: &  18.167: &          &         &  1                      \\
     5~~ &  OPH J171206.10-231818  &  258.02541  &  -23.30512  &   20.681  &  22.224  &          &         &                         \\
     6~~ &  OPH J171206.16-232218  &  258.02567  &  -23.37174  &   20.869  &  22.623  &  20.445  &         &                         \\
     7~~ &  OPH J171206.17-232251  &  258.02570  &  -23.38082  &   19.684: &  20.981: &  19.744: &         &  3                      \\
     8*  &  OPH J171207.00-232405  &  258.02918  &  -23.40148  &   16.152  &  17.233  &  15.211  &         & 2MASX J17120697-2324053 \\ 
     9~~ &  OPH J171207.25-232501  &  258.03020  &  -23.41707  &   20.220  &  21.707  &  19.951  &         &                         \\
    10~~ &  OPH J171207.33-232532  &  258.03055  &  -23.42559  &   19.898  &  21.124  &  19.909  &         &                         \\
    11~~ &  OPH J171207.41-231826  &  258.03088  &  -23.30737  &   21.409  &  22.736  &          &    1    &                         \\
    12~~ &  OPH J171207.45-232030  &  258.03106  &  -23.34188  &   16.375  &  17.400  &  15.606  &         &                         \\
    13~~ &  OPH J171207.62-231809  &  258.03174  &  -23.30255  &   23.017  &  23.783  &          &    3    &                         \\
    14~~ &  OPH J171208.13-231835  &  258.03386  &  -23.30983  &   20.526  &  21.697  &          &         &  1                      \\
    15~~ &  OPH J171208.31-232255  &  258.03461  &  -23.38196  &   20.786: &  21.487: &  20.585: &    3    &  3                      \\
    16~~ &  OPH J171208.39-232606  &  258.03497  &  -23.43500  &   18.951  &  19.823  &  18.559: &         & star near nucleus     \\
    17~~ &  OPH J171208.50-231908  &  258.03543  &  -23.31901  &   18.087  &  19.047  &  17.472  &         &                         \\
    18~~ &  OPH J171208.68-232236  &  258.03615  &  -23.37671  &   22.142  &  23.924::&  21.255  &    1    &                         \\
    19~~ &  OPH J171208.74-232219  &  258.03641  &  -23.37204  &   19.718  &  20.644  &  19.131  &    3    &                         \\
    20~~ &  OPH J171208.94-231801  &  258.03725  &  -23.30048  &   22.662  &  23.271  &          &    3    &                         \\
    21~~ &  OPH J171209.04-231808  &  258.03767  &  -23.30242  &   20.475  &  21.588  &          &         &                         \\
    22~~ &  OPH J171209.78-232020  &  258.04073  &  -23.33899  &   22.828  &  22.887: &  22.576  &    2    &                         \\
    23~~ &  OPH J171209.98-231845  &  258.04158  &  -23.31256  &   20.665  &  21.627  &          &         &                         \\
    24~~ &  OPH J171210.04-231753  &  258.04184  &  -23.29820  &   18.946  &  19.868  &          &         &                         \\
    25~~ &  OPH J171210.12-232445  &  258.04218  &  -23.41262  &   21.576  &  22.824  &  21.703: &    1    &                         \\
    26~~ &  OPH J171210.37-231756  &  258.04320  &  -23.29896  &   18.773: &  19.377: &          &         &  2                      \\
    27~~ &  OPH J171210.90-231754  &  258.04541  &  -23.29859  &   20.444::&  20.530::&          &    1    &  2                      \\
    28~~ &  OPH J171211.08-232251  &  258.04615  &  -23.38083  &   22.161  &  22.964  &  21.349  &    2    &                         \\
    29~~ &  OPH J171211.12-232504  &  258.04635  &  -23.41799  &   21.450  &  22.410  &  20.876  &    2    &                         \\
    30~~ &  OPH J171211.13-232604  &  258.04636  &  -23.43456  &   22.321  &  22.930  &  22.203: &    3    &                         \\
    31~~ &  OPH J171211.15-232415  &  258.04647  &  -23.40442  &   19.712  &  20.703  &  18.815  &    1    &  1                      \\
    32~~ &  OPH J171211.18-232500  &  258.04657  &  -23.41668  &   20.694  &  21.577  &  20.467  &         &                         \\
    33~~ &  OPH J171211.23-232006  &  258.04679  &  -23.33513  &   21.661  &  22.713  &  21.661: &    3    &                         \\
    34~~ &  OPH J171211.50-231821  &  258.04793  &  -23.30600  &   20.681  &  21.640  &          &         &                         \\
    35~~ &  OPH J171211.52-232557  &  258.04802  &  -23.43265  &   18.422  &  19.590  &  17.658  &         &                         \\
    36~~ &  OPH J171212.06-232434  &  258.05026  &  -23.40950  &   21.075  &  22.306  &  20.648  &    3    &  1                      \\
    37~~ &  OPH J171212.33-231824  &  258.05139  &  -23.30683  &   21.170  &  22.007  &          &    1    &                         \\
    38~~ &  OPH J171212.35-231743  &  258.05147  &  -23.29536  &   17.913  &  18.902  &          &         &                         \\
    39~~ &  OPH J171212.43-232158  &  258.05179  &  -23.36620  &   22.110  &  23.415  &  20.714  &    3    &                         \\
    40~~ &  OPH J171212.51-232119  &  258.05213  &  -23.35536  &   23.728  &  24.415  &  22.664  &    3    &                         \\
    41*  &  OPH J171212.61-232501  &  258.05253  &  -23.41716  &   20.396  &  21.297  &  19.761  &         &                         \\
    42~~ &  OPH J171212.76-232546  &  258.05317  &  -23.42969  &   21.266  &  22.060  &  20.540  &    2    &                         \\
    43~~ &  OPH J171214.15-231911  &  258.05897  &  -23.31989  &   22.734  &  23.369  &  22.509  &         &                         \\
    44~~ &  OPH J171215.13-232322  &  258.06306  &  -23.38954  &   19.280  &  20.475  &  18.915  &         &                         \\
    45~~ &  OPH J171215.24-232159  &  258.06352  &  -23.36655  &   20.758  &  21.915  &  20.186  &         &                         \\
    46~~ &  OPH J171215.60-231833  &  258.06501  &  -23.30928  &   18.949  &  19.948  &          &         &                         \\
    47~~ &  OPH J171216.10-231806  &  258.06707  &  -23.30168  &   22.428  &  23.008  &          &    3    &                         \\
    48*  &  OPH J171216.71-231931  &  258.06963  &  -23.32539  &   17.178  &  18.196  &  16.424  &         &                         \\
    49~~ &  OPH J171216.92-231946  &  258.07048  &  -23.32957  &   21.802  &  22.882  &  21.017  &         &                         \\
    50*  &  OPH J171217.22-232537  &  258.07174  &  -23.42704  &   17.895  &  18.757  &  17.333  &         &                         \\
    51~~ &  OPH J171217.56-232550  &  258.07316  &  -23.43066  &   20.851: &  21.164  &  19.774  &         &  1                      \\
    52~~ &  OPH J171217.95-232440  &  258.07479  &  -23.41120  &   22.095  &  23.309  &  21.734  &    1    &                         \\
    53~~ &  OPH J171218.18-232455  &  258.07574  &  -23.41533  &   22.101  &  23.655  &  21.003::&    3    &                         \\
    54~~ &  OPH J171218.50-232042  &  258.07709  &  -23.34512  &   17.474  &  18.573  &  16.740  &         &                         \\
    55~~ &  OPH J171218.51-231914  &  258.07714  &  -23.32057  &   22.212  &  23.398  &  21.227  &    1    &                         \\
    56~~ &  OPH J171218.71-231836  &  258.07794  &  -23.31010  &   21.227  &  22.740  &          &    1    &                         \\
    57~~ &  OPH J171218.82-232525  &  258.07840  &  -23.42361  &   21.418  &  22.879  &  20.883  &    2    &                         \\
    58*  &  OPH J171218.97-231921  &  258.07904  &  -23.32266  &   17.932  &  18.992  &  17.238  &         &                         \\
    59*  &  OPH J171218.98-232219  &  258.07910  &  -23.37198  &   15.541  &  16.654  &  14.614  &         & 2MASX J17121895-2322192 \\
    60~~ &  OPH J171219.27-232046  &  258.08030  &  -23.34625  &   21.602  &  23.237  &  20.777  &    3    &                         \\
         &                           &             &             &           &          &          &         &                         \\
\hline
\end{tabular}
\label{tab:photo_small}
\end{table*}

\begin{table*}[ht!]
\centering
\setcounter{table}{0}
\caption{Continued.}
\begin{tabular}{rllllllll}
\hline
\hline
         &                           &             &             &           &          &          &         &                         \\
Running  & OPH                       &  RA         &  DEC        &    $r'$   &    $g'$  &    $z'$  & Quality & Note                    \\
number   & name                      &  (J2000.0)  &  (J2000.0)  &           &          &          &         &                         \\
         &                           &             &             &           &          &          &         &                         \\
\hline
         &                           &             &             &           &          &          &         &                         \\
    61~~ &  OPH J171219.30-232012  &  258.08041  &  -23.33684  &   22.368  &  23.115  &  20.919: &    2    &                         \\
    62~~ &  OPH J171219.32-232059  &  258.08051  &  -23.34981  &   22.621  &  23.217  &  21.870  &    1    &                         \\
    63*  &  OPH J171219.91-232418  &  258.08294  &  -23.40510  &   15.405  &  16.560  &  14.440  &         & 2MASX J17121989-2324187 \\
    64~~ &  OPH J171220.07-232001  &  258.08362  &  -23.33384  &   21.770  &  22.375  &  21.125  &    3    &                         \\
    65*  &  OPH J171220.15-232343  &  258.08397  &  -23.39538  &   18.032  &  19.042  &  17.288  &         &                         \\
    66~~ &  OPH J171220.37-232551  &  258.08487  &  -23.43096  &   21.449  &  22.682  &  20.905  &         &                         \\
    67~~ &  OPH J171220.41-232124  &  258.08504  &  -23.35678  &   18.259  &  19.366  &  17.575  &         &  2                      \\
    68*  &  OPH J171220.68-232054  &  258.08615  &  -23.34832  &   18.458  &  19.561  &  17.738  &         &                         \\
    69*  &  OPH J171220.70-231831  &  258.08626  &  -23.30878  &   15.386  &  16.540  &          &         & 2MASX J17122070-2318318 \\
    70~~ &  OPH J171221.47-231745  &  258.08947  &  -23.29602  &   21.317  &  22.336  &          &    1    &   semi-stellar          \\
    71~~ &  OPH J171221.47-231954  &  258.08947  &  -23.33189  &   21.287  &  22.693  &  20.402  &         &                         \\
    72~~ &  OPH J171221.55-232305  &  258.08979  &  -23.38498  &   22.808  &  23.554  &  22.316  &    3    &                         \\
    73*  &  OPH J171221.60-232528  &  258.08999  &  -23.42448  &   18.650  &  19.600  &  18.003  &         &                         \\
    74*  &  OPH J171222.14-232604  &  258.09225  &  -23.43459  &   15.143  &  16.225  &  14.331  &         & 2MASX J17122214-2326048 \\
    75*  &  OPH J171223.04-232252  &  258.09598  &  -23.38113  &   16.654  &  17.762  &  15.900  &         & 2MASX J17122302-2322518 \\ 
    76*  &  OPH J171223.09-232156  &  258.09622  &  -23.36565  &   20.783  &  21.699  &  20.288  &         &                         \\
    77~~ &  OPH J171223.10-232112  &  258.09623  &  -23.35345  &   19.684  &  20.660  &  19.060  &         &  1                      \\
    78~~ &  OPH J171223.42-232053  &  258.09760  &  -23.34826  &   22.270  &  23.472  &  21.976  &    1    &                         \\
    79~~ &  OPH J171223.68-231813  &  258.09867  &  -23.30384  &   18.663  &  19.526  &          &         &                         \\
    80~~ &  OPH J171223.70-232313  &  258.09875  &  -23.38697  &   21.340  &  22.164  &  20.505  &    3    &                         \\
    81~~ &  OPH J171224.18-231828  &  258.10075  &  -23.30798  &   17.898  &  18.991  &          &         &                         \\
    82~~ &  OPH J171224.73-231821  &  258.10304  &  -23.30586  &   19.642  &  20.308  &          &         &                         \\
    83*  &  OPH J171225.43-232146  &  258.10595  &  -23.36299  &   17.743  &  18.881  &  16.959  &         &                         \\
    84~~ &  OPH J171226.61-232511  &  258.11089  &  -23.41990  &   22.098  &  23.544  &  21.323  &    3    &                         \\
    85~~ &  OPH J171226.62-232425  &  258.11093  &  -23.40713  &   21.460::&  21.194::&  20.744::&    3    &  2                      \\
    86~~ &  OPH J171226.79-232236  &  258.11164  &  -23.37674  &   19.481  &  20.768  &  18.580  &         &                         \\
    87*  &  OPH J171226.83-232249  &  258.11181  &  -23.38038  &   17.826  &  18.952  &  17.082  &         &                         \\
    88*  &  OPH J171226.86-234327  &  258.11194  &  -23.72424  &   16.621::&          &          &         &                         \\ 
    89~~ &  OPH J171227.25-232453  &  258.11354  &  -23.41477  &   19.787  &  20.502  &  19.135  &         &                         \\
    90~~ &  OPH J171227.36-232515  &  258.11402  &  -23.42091  &   22.263: &  22.945: &  22.135: &    2    &                         \\
    91*  &  OPH J171227.36-232225  &  258.11402  &  -23.37373  &   19.602::&  20.350::&          &         & superposed on cD galaxy \\
    92*  &  OPH J171227.72-232211  &  258.11549  &  -23.36978  &   15.284: &  16.386: &  13.484  &         & 2MASX J17122774-2322108 cD \\ 
    93~~ &  OPH J171228.62-231930  &  258.11924  &  -23.32521  &   22.648  &  23.106  &  21.349  &         &                         \\
    94~~ &  OPH J171228.66-231959  &  258.11942  &  -23.33316  &   21.902  &  23.142  &  21.067  &    2    &                         \\
    95~~ &  OPH J171229.14-232419  &  258.12141  &  -23.40546  &   19.465  &  20.438  &  18.859  &         &                         \\
    96~~ &  OPH J171229.39-231920  &  258.12246  &  -23.32233  &   19.199  &  20.259  &  17.880: &         &  1                      \\
    97*  &  OPH J171229.40-232241  &  258.12249  &  -23.37805  &   18.194  &  19.219  &  17.348  &         &                         \\
    98~~ &  OPH J171229.48-232123  &  258.12285  &  -23.35654  &   18.212  &  19.614  &  17.279  &         &                         \\
    99~~ &  OPH J171229.94-232632  &  258.12476  &  -23.44238  &   19.368  &  20.290  &          &         &                         \\
   100~~ &  OPH J171230.29-232604  &  258.12619  &  -23.43444  &   19.340  &  20.410  &  18.641  &         &                         \\
   101~~ &  OPH J171230.44-232057  &  258.12682  &  -23.34921  &   21.873  &  22.473  &  21.449  &    3    &                         \\
   102~~ &  OPH J171230.92-231850  &  258.12885  &  -23.31406  &   22.081  &  23.493  &          &    3    &                         \\
   103~~ &  OPH J171230.96-232320  &  258.12899  &  -23.38910  &   20.748  &  21.906  &  19.986  &         &                         \\
   104~~ &  OPH J171231.02-232448  &  258.12927  &  -23.41346  &   23.554: &  23.880: &  23.204: &    3    &                         \\
   105~~ &  OPH J171231.13-232211  &  258.12971  &  -23.36971  &   20.390  &  21.431  &  19.626  &    1    &  semi-stellar           \\
   106*  &  OPH J171231.55-232324  &  258.13146  &  -23.38999  &   17.959  &  19.006  &  17.156  &         &                         \\
   107~~ &  OPH J171231.81-231742  &  258.13254  &  -23.29525  &   22.789  &  23.471  &          &         &                         \\
   108~~ &  OPH J171231.97-232043  &  258.13319  &  -23.34550  &   18.483  &  19.721  &  17.662  &         &                         \\
   109~~ &  OPH J171232.03-232529  &  258.13345  &  -23.42494  &   22.936  &  23.699  &  22.509  &    3    &                         \\
   110~~ &  OPH J171233.25-231841  &  258.13854  &  -23.31157  &   19.647  &  20.375  &          &         &                         \\
   111~~ &  OPH J171233.36-232203  &  258.13898  &  -23.36771  &   21.705  &  23.032  &  21.282  &         &                         \\
   112~~ &  OPH J171233.54-232120  &  258.13973  &  -23.35579  &   19.366  &  20.466  &  18.465  &    1    &  semi-stellar           \\
   113~~ &  OPH J171234.01-232159  &  258.14170  &  -23.36643  &   23.363::&  24.694::&  21.790::&    3    &                         \\
   114~~ &  OPH J171235.46-231726  &  258.14777  &  -23.29077  &   21.386  &  22.427  &          &         &                         \\
   115~~ &  OPH J171235.64-232449  &  258.14852  &  -23.41374  &   20.806  &  21.617  &  19.896  &         &                         \\
   116~~ &  OPH J171235.74-232624  &  258.14892  &  -23.43999  &   19.592  &  20.784  &          &         &                         \\
   117*  &  OPH J171235.80-231920  &  258.14916  &  -23.32229  &   15.684  &  16.718  &  14.852  &         & 2MASX J17123580-2319208 \\
   118~~ &  OPH J171235.86-232409  &  258.14943  &  -23.40257  &   21.983  &  23.471  &  21.451  &    1    &                         \\
   119~~ &  OPH J171235.91-231954  &  258.14964  &  -23.33191  &   22.855  &  23.466  &  22.287  &    3    &                         \\
   120~~ &  OPH J171236.01-232335  &  258.15006  &  -23.39318  &   21.061  &  22.067  &  20.363  &         &                         \\
         &                           &             &             &           &          &          &         &                         \\
\hline
\end{tabular}
\end{table*}

\begin{table*}[ht!]
\centering
\setcounter{table}{0}
\caption{Continued.}
\begin{tabular}{rllllllll}
\hline
\hline
         &                           &             &             &           &          &          &         &                         \\
Running  & OPH                       &  RA         &  DEC        &    $r'$   &    $g'$  &    $z'$  & Quality & Note                    \\
number   & name                      &  (J2000.0)  &  (J2000.0)  &           &          &          &         &                         \\
         &                           &             &             &           &          &          &         &                         \\
\hline
         &                           &             &             &           &          &          &         &                         \\
   121~~ &  OPH J171236.21-232255  &  258.15089  &  -23.38199  &   22.620  &  24.090  &  22.120  &    3    &                         \\
   122~~ &  OPH J171236.31-232632  &  258.15128  &  -23.44226  &   20.866  &  21.679  &          &         &                         \\
   123*  &  OPH J171236.39-232114  &  258.15163  &  -23.35395  &   15.727  &  16.740  &  14.981  &         & 2MASX J17123638-2321138 \\
   124*  &  OPH J171236.61-232056  &  258.15254  &  -23.34913  &   19.605  &  20.505  &  18.912  &         &                         \\
   125~~ &  OPH J171237.09-232412  &  258.15454  &  -23.40351  &   17.959  &  19.011  &  17.230: &         & at field edge in z-band \\
   126~~ &  OPH J171237.32-231737  &  258.15548  &  -23.29364  &   19.779  &  20.597  &          &         &  2                      \\
   127~~ &  OPH J171237.37-231753  &  258.15572  &  -23.29807  &   22.910  &  25.070: &          &    1    &                         \\
   128~~ &  OPH J171237.55-232520  &  258.15647  &  -23.42226  &   21.172  &  21.882  &          &         &                         \\
   129~~ &  OPH J171237.64-232623  &  258.15683  &  -23.43980  &   15.671::&          &          &         &  2                      \\
   130~~ &  OPH J171237.65-232333  &  258.15686  &  -23.39253  &   22.128  &  23.306  &          &    3    &                         \\
   131~~ &  OPH J171237.95-232230  &  258.15811  &  -23.37511  &   18.365  &  19.292  &          &         &                         \\
   132~~ &  OPH J171238.19-232029  &  258.15914  &  -23.34154  &   19.642  &  21.057  &          &         &                         \\
   133~~ &  OPH J171238.41-232133  &  258.16004  &  -23.35923  &   22.639  &  23.895  &          &    3    &                         \\
   134~~ &  OPH J171238.81-232220  &  258.16171  &  -23.37238  &   19.764  &  21.090  &          &         &                         \\
   135~~ &  OPH J171239.26-232430  &  258.16358  &  -23.40850  &   19.010  &  19.822  &          &         &                         \\
   136~~ &  OPH J171239.38-232224  &  258.16410  &  -23.37359  &   20.166  &  22.034  &          &         &                         \\
   137~~ &  OPH J171239.42-232240  &  258.16423  &  -23.37790  &   21.170  &  22.796  &          &         &                         \\
   138~~ &  OPH J171239.83-232005  &  258.16597  &  -23.33486  &   19.438  &  20.168  &          &         &                         \\
   139~~ &  OPH J171240.15-232505  &  258.16730  &  -23.41831  &   16.299  &  17.354  &          &         &                         \\
   140~~ &  OPH J171241.01-232639  &  258.17087  &  -23.44419  &   15.6::  &          &          &         &  3                      \\   
   141~~ &  OPH J171241.03-232537  &  258.17097  &  -23.42717  &   18.288  &  19.563  &          &         &                         \\
   142~~ &  OPH J171241.44-232335  &  258.17265  &  -23.39326  &   19.296  &  20.464  &          &         &                         \\
   143~~ &  OPH J171241.56-231950  &  258.17315  &  -23.33065  &   20.428  &  21.657  &          &    1    &   semi-stellar          \\
   144~~ &  OPH J171241.73-232501  &  258.17387  &  -23.41709  &   20.760  &  21.885  &          &         &                         \\
   145~~ &  OPH J171242.25-232003  &  258.17605  &  -23.33418  &   16.067  &  17.101  &          &         &                         \\
   146~~ &  OPH J171242.31-232338  &  258.17629  &  -23.39414  &   22.810  &  24.163  &          &    1    &                         \\
   147*  &  OPH J171242.60-232417  &  258.17751  &  -23.40484  &   15.127  &  16.239  &          &         & 2MASX J17124256-2324167 \\
   148~~ &  OPH J171242.83-231938  &  258.17846  &  -23.32733  &   20.134: &  20.507: &          &         &  1                      \\
   149~~ &  OPH J171243.02-232452  &  258.17926  &  -23.41457  &   22.558  &  23.558  &          &    3    &                         \\
   150~~ &  OPH J171243.33-232558  &  258.18054  &  -23.43303  &   19.637  &  20.421  &          &         &                         \\
   151~~ &  OPH J171243.61-232125  &  258.18170  &  -23.35716  &   20.077  &  21.085  &          &         &                         \\
   152~~ &  OPH J171243.64-232038  &  258.18185  &  -23.34398  &   22.171  &  23.214  &          &    3    &                         \\
   153~~ &  OPH J171243.78-232541  &  258.18242  &  -23.42828  &   20.517: &  22.335: &          &    3    &                         \\
   154~~ &  OPH J171244.00-232056  &  258.18334  &  -23.34908  &   22.043  &  23.059  &          &         &                         \\
   155~~ &  OPH J171244.23-232124  &  258.18428  &  -23.35667  &   19.042  &  20.202  &          &         &                         \\
   156~~ &  OPH J171244.24-232300  &  258.18434  &  -23.38354  &   20.446  &  21.992  &          &         &                         \\
   157~~ &  OPH J171245.51-231900  &  258.18961  &  -23.31673  &   21.310  &  22.130  &          &    1    &                         \\
   158~~ &  OPH J171245.61-232019  &  258.19004  &  -23.33875  &   17.504  &  18.478  &          &         &                         \\
   159~~ &  OPH J171246.11-231819  &  258.19212  &  -23.30538  &   21.027  &  21.930  &          &    3    &                         \\
   160~~ &  OPH J171246.33-231738  &  258.19306  &  -23.29391  &   20.302  &  21.258  &          &         &                         \\
   161~~ &  OPH J171247.16-231854  &  258.19650  &  -23.31515  &   20.575  &  21.300  &          &         &                         \\
   162~~ &  OPH J171247.68-232127  &  258.19865  &  -23.35757  &   23.125  &  23.644  &          &    3    &                         \\
         &                           &             &             &           &          &          &         &                         \\
\hline
\end{tabular}
\end{table*}

\begin{table*}[ht!]
\centering
\setcounter{table}{1}
\caption{Objects with a measured spectroscopic redshift but outside the VLT/FORS2 area.
The columns are: (1) running number, (2) object name, (3) and (4) RA and Dec (J2000.0),
(5) $r'$ band magnitude, (6) other identification.}
\begin{tabular}{rlllll}
\hline
\hline
       &                         &             &             &           &                          \\
Running & OPH                    & RA        & DEC           & $r'$   & other \\
number  & name                   & (J2000.0) & (J2000.0)     &        & name   \\
       &                         &             &             &           &                          \\
\hline
       &                         &             &             &           &                          \\
     1 &  OPH J171014.78-233404  &  257.56158  &  -23.56784  &   15.672 &  2MASX J17101464-2333587 \\
     2 &  OPH J171030.94-233839  &  257.62893  &  -23.64431  &   15.886 &  2MASX J17103094-2338399 \\
     3 &  OPH J171035.72-232925  &  257.64883  &  -23.49036  &   16.611 &  2MASX J17103573-2329248 \\
     4 &  OPH J171040.67-232422  &  257.66945  &  -23.40633  &   16.179 &  2MASX J17104066-2324228 \\
     5 &  OPH J171041.87-231335  &  257.67446  &  -23.22643  &   16.308 &  2MASX J17104188-2313348 \\
     6 &  OPH J171044.00-231400  &  257.68333  &  -23.23336  &   16.284 &  2MASX J17104398-2313598 \\
     7 &  OPH J171045.89-231923  &  257.69121  &  -23.32331  &   16.067 &  2MASX J17104588-2319238 \\
     8 &  OPH J171049.53-233624  &  257.70638  &  -23.60683  &   15.982 &  2MASX J17104954-2336246 \\
     9 &  OPH J171058.45-225927  &  257.74355  &  -22.99086  &   15.660 &  2MASX J17105842-2259266 \\
    10 &  OPH J171101.25-231107  &  257.75522  &  -23.18542  &   16.108 &                          \\     
    11 &  OPH J171121.22-231849  &  257.83843  &  -23.31380  &   14.801 &  2MASX J17112119-2318497 \\
    12 &  OPH J171122.02-230529  &  257.84176  &  -23.09154  &   16.834 &  2MASX J17112206-2305297 \\
    13 &  OPH J171124.03-230042  &  257.85012  &  -23.01172  &   15.968 &  2MASX J17112403-2300427 \\
    14 &  OPH J171125.89-231851  &  257.85789  &  -23.31420  &   16.174 &  2MASX J17112592-2318507 \\
    15 &  OPH J171127.15-230922  &  257.86313  &  -23.15616  &   16.399 &  6dF J1711272-230922     \\
    16 &  OPH J171127.20-232309  &  257.86332  &  -23.38601  &   16.263 &  2MASX J17112722-2323097 \\
    17 &  OPH J171128.02-234420  &  257.86677  &  -23.73902  &   16.333 &  2MASX J17112802-2344197 \\
    18 &  OPH J171136.00-230306  &  257.89998  &  -23.05172  &   15.595 &  2MASX J17113598-2303057 \\
    19 &  OPH J171137.53-225846  &  257.90636  &  -22.97946  &   15.445 &  2MASX J17113750-2258457 \\
    20 &  OPH J171143.43-233106  &  257.93097  &  -23.51835  &   15.718 &  2MASX J17114344-2331057 \\
    21 &  OPH J171144.13-230339  &  257.93386  &  -23.06101  &   15.082 &  2MASX J17114409-2303397 \\
    22 &  OPH J171145.31-230844  &  257.93881  &  -23.14556  &   15.574 &  2MASX J17114534-2308447 \\
    23 &  OPH J171151.29-230136  &  257.96372  &  -23.02688  &   15.080 &  2MASX J17115130-2301363 \\
    24 &  OPH J171155.39-230942  &  257.98079  &  -23.16187  &   15.242 &  2MASX J17115542-2309423 \\
    25 &  OPH J171156.54-231501  &  257.98557  &  -23.25032  &   15.682 &  2MASX J17115651-2315013 \\
    26 &  OPH J171156.63-230250  &  257.98594  &  -23.04726  &   16.319 &  2MASX J17115666-2302503 \\
    27 &  OPH J171157.23-230211  &  257.98846  &  -23.03648  &   15.201 &  2MASX J17115724-2302113 \\
    28 &  OPH J171157.46-232938  &  257.98941  &  -23.49403  &   16.537 &  2MASXi J1711574-232938  \\
    29 &  OPH J171157.70-234207  &  257.99041  &  -23.70216  &   16.380 &  2MASX J17115766-2342073 \\
    30 &  OPH J171209.06-232826  &  258.03774  &  -23.47404  &   16.078 &  2MASX J17120908-2328263 \\
    31 &  OPH J171210.92-233300  &  258.04550  &  -23.55018  &   16.513 &  2MASX J17121090-2333003 \\
    32 &  OPH J171217.94-230342  &  258.07475  &  -23.06182  &   16.875 &  2MASX J17121796-2303422 \\
    33 &  OPH J171219.71-230828  &  258.08212  &  -23.14121  &   16.239 &  2MASX J17121970-2308288 \\
    34 &  OPH J171219.79-231541  &  258.08247  &  -23.26150  &   15.288 &  2MASX J17121976-2315418 \\
    35 &  OPH J171220.86-231004  &  258.08691  &  -23.16801  &   15.332 &  2MASX J17122085-2310048 \\
    36 &  OPH J171220.94-231610  &  258.08727  &  -23.26970  &   16.492 &  2MASX J17122092-2316108 \\
    37 &  OPH J171221.13-232957  &  258.08805  &  -23.49916  &   15.290 &  2MASX J17122113-2329568 \\
    38 &  OPH J171223.90-225159  &  258.09960  &  -22.86646  &   16.598 &  2MASXiJ1712239 -225159  \\
    39 &  OPH J171227.78-230811  &  258.11574  &  -23.13658  &   16.431 &  2MASX J17122782-2308108 \\
    40 &  OPH J171228.80-230803  &  258.11998  &  -23.13422  &   16.183 &  2MASX J17122884-2308038 \\
    41 &  OPH J171233.29-230734  &  258.13870  &  -23.12631  &   15.720 &  2MASX J17123325-2307348 \\
    42 &  OPH J171238.46-233145  &  258.16023  &  -23.52930  &   15.784 &  2MASX J17123843-2331458 \\
    43 &  OPH J171240.94-232942  &  258.17060  &  -23.49512  &   16.140 &  2MASX J17124097-2329427 \\
    44 &  OPH J171247.65-230151  &  258.19854  &  -23.03087  &   16.905 &  2MASX J17124767-2301517 \\
    45 &  OPH J171249.23-234834  &  258.20511  &  -23.80948  &   17.600 &                          \\
    46 &  OPH J171249.58-230720  &  258.20657  &  -23.12233  &   15.373 &  2MASX J17124957-2307207 \\
    47 &  OPH J171253.26-231705  &  258.22191  &  -23.28487  &   16.724 &                          \\
    48 &  OPH J171255.07-225700  &  258.22945  &  -22.95003  &   15.507 &  2MASX J17125506-2257000 \\
    49 &  OPH J171306.82-225535  &  258.27843  &  -22.92641  &   15.904 &  2MASX J17130678-2255350 \\
    50 &  OPH J171317.53-233429  &  258.32303  &  -23.57474  &   15.152 &  2MASX J17131749-2334290 \\
    51 &  OPH J171323.91-233450  &  258.34963  &  -23.58056  &   16.369 &  2MASX J17132392-2334506 \\
    52 &  OPH J171324.19-233914  &  258.35081  &  -23.65413  &   16.105 &  2MASX J17132420-2339146 \\
    53 &  OPH J171326.45-233201  &  258.36020  &  -23.53382  &   16.095 &  2MASX J17132646-2332016 \\
    54 &  OPH J171328.44-230840  &  258.36852  &  -23.14461  &   16.167 &  2MASX J17132844-2308406 \\
    55 &  OPH J171334.02-233733  &  258.39177  &  -23.62591  &   16.105 &  2MASX J17133402-2337336 \\
    56 &  OPH J171339.33-231814  &  258.41387  &  -23.30412  &   15.599 &                          \\
    57 &  OPH J171351.94-230328  &  258.46642  &  -23.05802  &   16.522 &  2MASX J17135195-2303286 \\
    58 &  OPH J171352.23-232155  &  258.46763  &  -23.36535  &   15.925 &  2MASX J17135221-2321556 \\
    59 &  OPH J171354.89-231618  &  258.47871  &  -23.27181  &   16.110 &  2MASX J17135490-2316186 \\
    60 &  OPH J171358.48-225156  &  258.49365  &  -22.86564  &   15.859 &  2MASX J17135847-2251566 \\
    61 &  OPH J171401.72-231302  &  258.50717  &  -23.21740  &   15.035 &  2MASX J17140172-2313026 \\
    62 &  OPH J171409.76-233153  &  258.54066  &  -23.53157  &   15.028 &  2MASX J17140971-2331536 \\
    63 &  OPH J171422.04-230018  &  258.59185  &  -23.00507  &   16.762 &  2MASX J17142206-2300182 \\
    64 &  OPH J171423.86-231048  &  258.59942  &  -23.18012  &   15.653 &  2MASX J17142386-2310482 \\ 
    65 &  OPH J171424.83-231347  &  258.60345  &  -23.22983  &   14.735 &  2MASX J17142479-2313472 \\
       &                         &             &             &           &                         \\
\hline
\end{tabular}
\label{tab:photo_r_ext}
\end{table*}

\begin{table*} 
\centering
\setcounter{table}{2}
\caption{Spectroscopic redshift catalogue. 
  The columns are: (1)~running number (for objects with *, photometric data are given
  in Tables A1 (VLT/FORS2 area) and A2 (outside the VLT/FORS2 area), (2) object name,
  (3)~spectroscopic redshift, (4)~reference to the origin of the redshift: 1 for 
 our FORS2 data, 2 for our 6df, CTIO 1.5m, and Lick 3m data (Wakamatsu et al. 2005),
 and 3 for NED.  }
\begin{tabular}{rlll}
\hline
\hline
       &                         &       &  \\
Number & name     & redshift   & reference \\
       &                         &       &  \\
\hline
       &                         &       &  \\
  1~~  & 2MASX J17071767-2406391 & 0.0332&3  \\
  2~~  & 2MASX J17073583-2353321 & 0.0317&3  \\
  3~~  & 2MASX J17075426-2346173 & 0.0295&3  \\
  4~~  & 2MASX J17075972-2346163 & 0.0320&3  \\
  5~~  & 2MASX J17080961-2232072 & 0.0290&3  \\
  6~~  & 2MASX J17081885-2252587 & 0.0259&3  \\
  7~~  & 2MASX J17082726-2326327 & 0.0327&2  \\
  8~~  & 2MASX J17082979-2256397 & 0.0258&2  \\
  9~~  & 2MASX J17083142-2217447 & 0.0296&3  \\
 10~~  & 2MASX J17084712-2329350 & 0.0275&2  \\
 11~~  & 2MASX J17085724-2346260 & 0.0269&2  \\
 12~~  & 2MASX J17090361-2327449 & 0.0329&2  \\
 13~~  & 2MASX J17091029-2233179 & 0.0279&2  \\
 14~~  & 2MASX J17091341-2345468 & 0.0321&2  \\
 15~~  & 2MASX J17092831-2226505 & 0.0310&3  \\
 16~~  & 2MASX J17092925-2259155 & 0.0296&2  \\
 17~~  & 2MASX J17093322-2232395 & 0.0290&2  \\
 18~~  & 2MASX J17093341-2350245 & 0.0220&2  \\
 19~~  & 2MASX J17093394-2238225 & 0.0307&2  \\
 20~~  & 2MASX J17093813-2358058 & 0.0312&2  \\
 21~~  & 2MASX J17093810-2305055 & 0.0265&2  \\
 22~~  & 2MASX J17094318-2231395 & 0.0324&2  \\
 23~~  & 1RXS  J170944.9-234658  & 0.0364&2  \\
 24~~  & 2MASX J17095181-2313038 & 0.0323&2  \\
 25~~  & 2MASX J17095391-2250048 & 0.0293&2  \\
 26~~  & 2MASX J17095655-2416307 & 0.0313&2  \\
 27~~  & 2MASX J17100247-2316308 & 0.0277&2  \\
 28~~  & 2MASX J17100581-2315138 & 0.0285&2  \\
 29*   & 2MASX J17101464-2333587 & 0.0277&2  \\
 30~~  & 2MASX J17101930-2217218 & 0.0298&3  \\
 31*   & 2MASX J17103094-2338399 & 0.0339&2  \\
 32*   & 2MASX J17103573-2329248 & 0.0237&2  \\
 33*   & 2MASX J17104066-2324228 & 0.0265&2  \\
 34~~  & 2MASX J17104160-2411341 & 0.0371&2  \\
 35*   & 2MASX J17104188-2313348 & 0.0255&2  \\
 36*   & 2MASX J17104398-2313598 & 0.0325&2  \\
 37~~  & 2MASX J17104452-2359408 & 0.0279&2  \\ 
 38~~  & 2MASX J17104522-2247088 & 0.0357&2  \\
 39*   & 2MASX J17104588-2319238 & 0.0307&2  \\
 40*   & 2MASX J17104954-2336246 & 0.0311&2  \\
 41*   & 2MASX J17105842-2259266 & 0.0233&2  \\
 42*   & OPH J171101.25-231107   & 0.0259&2  \\
 43~~  & 2MASX J17111189-2354508 & 0.0266&2  \\
 44~~  & 2MASX J17112078-2234457 & 0.0322&2  \\
 45*   & 2MASX J17112119-2318497 & 0.0274&3  \\
 46*   & 2MASX J17112206-2305297 & 0.0302&2  \\
 47*   & 2MASX J17112403-2300427 & 0.0279&2  \\
 48*   & 2MASX J17112592-2318507 & 0.0325&2  \\
 49*   & 6dF   J1711272-230922   & 0.0369&2  \\
 50*   & 2MASX J17112722-2323097 & 0.0315&2  \\
 51~~  & 2MASX J17112794-2404348 & 0.0311&2  \\
 52*   & 2MASX J17112802-2344197 & 0.0294&2  \\
 53~~  & 2MASX J17113574-2413554 & 0.0317&2  \\
 54*   & 2MASX J17113598-2303057 & 0.0238&3  \\
 55*   & 2MASX J17113750-2258457 & 0.0288&2  \\
 56~~  & OPH J171139.58-235053   & 0.0322&2  \\
 57*   & 2MASX J17114344-2331057 & 0.0333&2  \\
 58*   & 2MASX J17114409-2303397 & 0.0274&3  \\
 59*   & 2MASX J17114534-2308447 & 0.0334&2  \\
 60*   & 2MASX J17115130-2301363 & 0.0331&3  \\
       &                         &       &  \\
\hline
\end{tabular}
\label{tab:redshifts}
\end{table*}

\begin{table*}[ht!]
\centering
\setcounter{table}{2}
\caption{Continued.}
\begin{tabular}{rlll}
\hline
\hline
     &                         &       &  \\
Number & name     & redshift   & reference \\
     &                         &       &  \\
\hline
     &                         &       &  \\
 61*  & 2MASX J17115542-2309423 & 0.0269&3 \\
 62*  & 2MASX J17115651-2315013 & 0.0322&2 \\
 63*  & 2MASX J17115666-2302503 & 0.0305&3 \\
 64*  & 2MASX J17115724-2302113 & 0.0326&3 \\
 65*  & 2MASXi J1711574-232938  & 0.0301&2 \\
 66*  & 2MASX J17115766-2342073 & 0.0340&2 \\
 67~~ & 2MASX J17115865-2438004 & 0.0297&3 \\
 68~~ & 2MASX J17115882-2359083 & 0.0347&2 \\
 69*  & 2MASX J17120697-2324053 & 0.0318&1 \\ 
 70*  & 2MASX J17120908-2328263 & 0.0249&2 \\
 71*  & 2MASX J17121090-2333003 & 0.0307&2 \\
 72~~ & 2MASX J17121160-2227513 & 0.0319&2 \\
 73*  & OPH J171212.61-232501   & 0.0306&1 \\ 
 74*  & OPH J171216.71-231931   & 0.0206&1 \\
 75*  & OPH J171217.22-232537   & 0.0243&1 \\
 76*  & 2MASX J17121796-2303422 & 0.0237&2 \\
 77*  & OPH J171218.97-231921   & 0.0322&1 \\
 78*  & 2MASX J17121895-2322192 & 0.0275&1 \\
 79*  & 2MASX J17121970-2308288 & 0.0276&2 \\
 80*  & 2MASX J17121976-2315418 & 0.0299&3 \\
 81*  & 2MASX J17121989-2324187 & 0.0296&2 \\
 82*  & OPH J171220.15-232343   & 0.0262&1 \\
 83*  & OPH J171220.68-232054   & 0.0311&1 \\
 84*  & 2MASX J17122070-2318318 & 0.0366&1 \\
 85*  & 2MASX J17122085-2310048 & 0.0356&2 \\
 86*  & 2MASX J17122092-2316108 & 0.0240&2 \\
 87*  & 2MASX J17122113-2329568 & 0.0306&3 \\
 88*  & OPH J171221.60-232528   & 0.0268&1 \\
 89*  & 2MASX J17122214-2326048 & 0.0264&1 \\
 90*  & 2MASX J17122302-2322518 & 0.0288&2 \\
 91*  & OPH J171223.09-232156   & 0.0276&1 \\
 92~~ & 2MASX J17122328-2231578 & 0.0318&2 \\
 93*  & 2MASXi J1712239-225159  & 0.0292&2 \\
 94*  & OPH J171225.43-232146   & 0.0283&1 \\
 95*  & OPH J171226.86-234327   & 0.0306&2 \\
 96*  & OPH J171226.83-232249   & 0.0373&1 \\ 
 97*  & OPH J171227.36-232225   & 0.0285&1 \\
 98*  & 2MASX J17122774-2322108 & 0.0295&2 \\
 99*  & 2MASX J17122782-2308108 & 0.0349&2 \\
100*  & 2MASX J17122884-2308038 & 0.0365&2 \\
101*  & OPH J171229.40-232241   & 0.0249&1 \\
102*  & OPH J171231.55-232324   & 0.0276&1 \\
103*  & 2MASX J17123325-2307348 & 0.0330&2 \\
104*  & 2MASX J17123580-2319208 & 0.0321&2 \\
105~~ & 2MASX J17123583-2222498 & 0.0279&3 \\
106*  & 2MASX J17123638-2321138 & 0.0356&1 \\
107*  & OPH J171236.61-232056   & 0.0281&1 \\
108*  & 2MASX J17123843-2331458 & 0.0262&2 \\
109*  & 2MASX J17124097-2329427 & 0.0346&2 \\
110*  & 2MASX J17124256-2324167 & 0.0288&3 \\
111~~ & 2MASX J17124278-2435477 & 0.0243&3 \\
112*  & OPH J171249.23-234834   & 0.0365&2 \\
113*  & 2MASX J17124767-2301517 & 0.0318&2 \\
114*  & 2MASX J17124957-2307207 & 0.0273&2 \\
115~~ & 2MASX J17125134-2415096 & 0.0282&3 \\
116*  & OPH J171253.26-231705   & 0.0260&2 \\
117*  & 2MASX J17125506-2257000 & 0.0323&2 \\
118~~ & 2MASX J17125738-2358196 & 0.0239&2 \\
119~~ & OPH J171301.84-223836   & 0.0317&2 \\
120~~ & 2MASX J17130287-2354020 & 0.0282&2 \\
      &                         &       &  \\
\hline
\end{tabular}
\end{table*}

\begin{table*}[ht!]
\centering
\setcounter{table}{2}
\caption{Continued.}
\begin{tabular}{rlll}
\hline
\hline
       &                         &       &  \\
Number & name     & redshift    & reference \\
       &                         &       &  \\
\hline
      &                         &       &  \\
121*  & 2MASX J17130678-2255350 & 0.0234&2 \\
122*  & 2MASX J17131749-2334290 & 0.0295&3 \\
123*  & 2MASX J17132392-2334506 & 0.0324&2 \\
124*  & 2MASX J17132420-2339146 & 0.0280&2 \\
125*  & 2MASX J17132646-2332016 & 0.0342&2 \\
126*  & 2MASX J17132844-2308406 & 0.0280&2 \\
127~~ & 2MASX J17132963-2234086 & 0.0323&2 \\
128*  & 2MASX J17133402-2337336 & 0.0252&2 \\
129~~ & 2MASX J17133908-2239286 & 0.0309&2 \\
130*  & OPH J171339.33-231814   & 0.0315&2 \\
131~~ & 2MASX J17134226-2353126 & 0.0297&2 \\
132~~ & 2MASX J17135024-2216106 & 0.0329&3 \\
133~~ & 2MASXiJ1713516-240757   & 0.0258&2  \\
134*  & 2MASX J17135195-2303286 & 0.0304&2  \\
135*  & 2MASX J17135221-2321556 & 0.0285&2  \\
136*  & 2MASX J17135490-2316186 & 0.0330&2  \\
137*  & 2MASX J17135847-2251566 & 0.0297&2  \\
138*  & 2MASX J17140172-2313026 & 0.0284&2  \\
139*  & 2MASX J17140971-2331536 & 0.0293&2  \\
140*  & 2MASX J17142206-2300182 & 0.0344&2  \\
141*  & 2MASX J17142386-2310482 & 0.0308&2  \\
142*  & 2MASX J17142479-2313472 & 0.0262&3  \\
143~~ & 2MASX J17143041-2408063 & 0.0256&2  \\
144~~ & 2MASX J17145813-2221299 & 0.0290&3  \\
145~~ & 2MASX J17150845-2354140 & 0.0303&2  \\
146~~ & 2MASX J17150925-2341020 & 0.0324&2  \\
147~~ & 2MASX J17151484-2326579 & 0.0330&2  \\
148~~ & 2MASX J17151503-2301319 & 0.0285&2  \\
149~~ & 2MASX J17154530-2340044 & 0.0284&2  \\
150~~ & 2MASX J17155885-2342043 & 0.0266&2  \\
151~~ & 2MASX J17165090-2324074 & 0.0286&3  \\
152~~ & 2MASX J17181000-2313584 & 0.0262&3  \\
      &                         &       &  \\
\hline
\end{tabular}
\end{table*}

\end{document}